\documentclass[12pt,dvipsnames]{article}

\usepackage{setspace}

\usepackage[utf8]{inputenc}
\usepackage{jheppub}
\usepackage{epsfig}
\usepackage{bbm}
\usepackage{bbding}
\usepackage{amsmath}
\usepackage{amsfonts}
\usepackage{amssymb}
\usepackage{mathtools}
\usepackage{dsfont}

\usepackage{graphicx}
\usepackage{longtable}
\usepackage{pdflscape}
\usepackage{subcaption}
\usepackage{psfrag}
\usepackage{hyperref}
 
\usepackage[capitalise]{cleveref}
\usepackage{tikz}
\usetikzlibrary{positioning,arrows.meta}
\usepackage{pgfplots}
\usepackage{pgfplotstable}
\usepackage{subfiles}
\usepackage{longtable}

\usepackage{tabularx}
\usepackage{ltablex}
\usepackage{booktabs}
\usepackage[export]{adjustbox}
\usepackage{listings}
\usepackage{multirow} 
\usepackage{cancel,slashed}
\usepackage{bbm}
\usepackage{graphicx}
\usepackage{latexsym}
\usepackage{transparent}
\usepackage{mathtools}
\usepackage{array}
\usepackage{makecell}
\usepackage{graphics,psfrag}
\usepackage{placeins}
\usepackage{nowidow}
\usepackage{listings}
\usepackage[normalem]{ulem}
\usepackage{environ}
\usepackage{enumerate}
\usetikzlibrary{positioning,arrows.meta}
\usetikzlibrary{positioning,trees,decorations.pathmorphing,decorations.markings,decorations.pathreplacing,calc,shapes,shapes.geometric,shapes.symbols,patterns,arrows}
\usetikzlibrary{fadings}
\usetikzlibrary{decorations.shapes}
\usepackage{amsmath}
\usepackage{amssymb}
\usepackage{amsthm}
\usepackage{soul}

\newcommand{\newshortstack}[1]
{\begingroup\renewcommand{\arraystretch}{1.1}
\ifmmode
\begin{array}{c}#1\end{array}%
\else
\begin{tabular}{c}#1\end{tabular}%
\fi
\endgroup}

\pgfplotsset{
        compat=1.9,
        compat/bar nodes=1.8,
    }

\theoremstyle{definition}

\newcommand{\be}{\begin{equation}}
	\newcommand{\ee}{\end{equation}}
\newcommand{\bea}{\begin{eqnarray}}
	\newcommand{\eea}{\end{eqnarray}}

\renewcommand{\epsilon}{\varepsilon}

\newcommand{\ben}{\begin{enumerate}}
	\newcommand{\een}{\end{enumerate}}
\newcommand{\bei}{\begin{itemize}}
	\newcommand{\eei}{\end{itemize}}
	
	\makeatletter
\tikzset{
    dot diameter/.store in=\dot@diameter,
    dot diameter=3pt,
    dot spacing/.store in=\dot@spacing,
    dot spacing=10pt,
    dots/.style={
        line width=\dot@diameter,
        line cap=round,
        dash pattern=on 0pt off \dot@spacing
    }
}
\makeatother
	
\tikzset{decorate sep/.style 2 args=
{decorate,decoration={shape backgrounds,shape=circle,shape size=#1,shape sep=#2}}}
\newtheorem*{theorem*}{Theorem}

\def\im{{\text{Im} \,}}

\def\tr{{\text{tr} \,}}

\renewcommand{\arraystretch}{1.2} 
\definecolor{NRG}{rgb}{0.172549, 0.627451, 0.172549}

\graphicspath{{Figures/}}

\preprint{DESY 25-113}

\title{
Towards a Heterotic Axiverse}
\author[a]{Jacob M. Leedom,}
\author[b,c]{Margherita Putti,}
\author[b]{$and$ Alexander Westphal}

\affiliation[a]{CEICO, Institute of Physics of the Czech Academy of Sciences, Na Slovance 2, 182 00 Prague 8, Czech Republic}
\affiliation[b]{ Deutsches Elektronen-Synchrotron DESY, Notkestr. 85, 22607 Hamburg, Germany}
\affiliation[c]{II. Institut f\"{u}r Theoretische Physik, Universit\"{a}t Hamburg
Luruper Chaussee 149, 22607 Hamburg, Germany}

\emailAdd{leedom@fzu.cz}
\emailAdd{margherita.putti@desy.de}
\emailAdd{alexander.westphal@desy.de}

\abstract{In this paper we initiate a broad study of some central properties of the string axiverse arising from Calabi-Yau compactifications of the perturbative heterotic $E_8\times E_8$ theory. Along this road toward a heterotic axiverse, we characterize the generic structure of the axion mass spectrum and the effective couplings of the non-QCD heterotic axions to Abelian and non-Abelian gauge fields and discuss their implications for cosmology, particle phenomenology, and the QCD axion quality problem. We also provide arguments that the heterotic axion masses are bounded from below much more strongly than, for example, the spectrum in type IIB compactifications.}

\begin{document}

\maketitle
\pagestyle{plain}

\section{Introduction}

Of the many potential windows into novel fundamental physics, axions remain one of the most promising. An axion that couples to the Chern-Simons (CS) term of Quantum Chromodynamics, dubbed the QCD axion~\cite{Weinberg:1977ma,Wilczek:1977pj}, can solve the Strong CP problem via the Peccei-Quinn mechanism~\cite{Peccei:1977ur,Peccei:1977hh} while also explaining the observed dark matter via the misalignment mechanism~\cite{Preskill:1982cy,Abbott:1982af,Dine:1982ah}.

Axions are equally well-motivated from the standpoint of string theory, where they arise in compactifications as zero modes from dimensional reduction of higher-form gauge fields. The perturbative shift symmetries of these string axions, related to the gauge symmetry of the higher-form fields, makes them promising candidates for both inflationary dynamics and dark matter. The number of axions in a string compactification can be quite large, giving rise to the so-called String Axiverse~\cite{Arvanitaki:2009fg,Cicoli:2012sz,Demirtas:2018akl,Gendler:2023kjt}.

Due to their plenitude and potentially light masses, axions may be a crucial bridge between the observable universe and string theory. To fully realize this potential, key aspects of string axions must be understood, such as their mass spectrum and coupling strength to Standard Model (SM) states. This direct coupling to the SM could then be exploited to uncover the string axiverse. However, it is certainly possible, if not likely, that many axions in the axiverse do not couple directly to the SM. In the absence of such couplings, one can still hope to detect string axions via gravitational signatures, such as tensor modes produced during inflation if the axion is the inflaton or is present as a spectator field~\cite{Anber:2009ua,Dimastrogiovanni:2012ew,Namba:2015gja,Peloso:2016gqs,DAmico:2021vka,DAmico:2021fhz,Dimastrogiovanni:2023juq}. For a review of the underlying production mechanism see e.g.~\cite{Pajer:2013fsa}.  

Axions have been extensively studied in the context of Type IIB orientifold compactifications~\cite{Cicoli:2012sz,Hebecker:2018yxs,Cicoli:2021gss,Carta:2021uwv,Demirtas:2021gsq,Gendler:2023kjt,Sheridan:2024vtt,Cheng:2025ggf}. IIB models offer many advantages since moduli stabilization is relatively well-understood and concrete model-building tools like flux compactifications, KKLT, and the Large Volume Scenario are readily available. Direct couplings of IIB string axions were studed in~\cite{Cicoli:2012sz,Gendler:2023kjt} while spectator axions and their CS couplings were examined in~\cite{Dimastrogiovanni:2023juq}. Furthermore, the QCD axion in type IIB and its potential quality problem where studied at length in~\cite{Demirtas:2021gsq}.

Despite the impressive tools available in type IIB compactifications, a major drawback of these models arise from realizing gauge theory sectors. As the type IIB perturbative spectrum lacks non-Abelian gauge theories, one must introduce non-perturbative objects such as D-branes that will furnish viable particle physics sectors. In contrast, heterotic string theories display the highly attractive feature of containing non-Abelian gauge fields even at the perturbative level, which facilitates the construction of GUTs and Standard Model-like spectra without the need for D-branes or localized sources. Axions also appear in heterotic compactifications, descending from the 10d NS-NS 2-form either as the universal, model-independent axion dual of the 4d 2-form or as model-dependent axions associated with the internal cohomology of the compactification manifold. However, a fully controlled cosmological setup, particularly one incorporating inflation and moduli stabilization, is still lacking in this framework. Recent progress, including mechanisms invoking NS-NS 3-form flux, gauge bundles, worldsheet instantons and gaugino condensation~\cite{deCarlos:1992kox,Gukov:2003cy,Gaillard:2007jr,Serone:2007sv,Anderson:2010mh,Dundee:2010sb,Parameswaran:2010ec,Cicoli:2013rwa,Leedom:2022zdm} as well as perturbative $\alpha'$, string loop, and non-perturbative corrections to the K\"ahler potential~\cite{Shenker:1990uf,Candelas:1990rm,Barreiro:1997rp,Becker:2002nn,Anguelova:2010ed}, suggests new possibilities for vacuum stabilization and cosmological model-building.

Despite the above attractive features, the heterotic sector of the axiverse is largely unexplored. Early studies of axions in heterotic theories focused on realizing the QCD axion and situations to avoid the quality problem~\cite{Choi:1985je,Choi:1985bz,Banks:1996ss,Gaillard:2005gj,Acharya:2010zx}. More recent works~\cite{Agrawal:2022lsp,Agrawal:2024ejr} as well as a forthcoming work~\cite{TimoReig} study the realization of a QCD axion in heterotic CY compactifications from a linear combination of the universal 4D axion and the NS-NS 2-form axions of heterotic string theory, establishing that with very few limited exceptions such a heterotic QCD axion will acquire values in the $g_{a\gamma\gamma}-m_a$-plane above the so-called QCD axion line. 

\begin{table}[t!]
\centering
\renewcommand{\arraystretch}{1.4}
\begin{tabular}{|c|c|c|c|c|}
\hline
\textbf{GC} &\textbf{$v_1\sim v_2$}& \textbf{Masses} & \textbf{CP} & \textbf{FDM}  \\
\hline
$\times$ & \checkmark & $m_{\varphi_2}^2\sim \Lambda_{\text{QCD}}^4  \ll m_{\varphi_1}^2\,,\, m_{\varphi_3}^2\sim\Lambda_{\text{ws}}^4$ & \checkmark  & $\times $\\
\hline
$\times$ &$\times$  & $m_{\varphi_1}^2\sim\epsilon \Lambda_{\text{ws}}^4 \ll m_{\varphi_2}^2\sim \Lambda_{\text{QCD}}^4\ll  m_{\varphi_3}^2\sim\Lambda_{\text{ws}}^4$ & \checkmark & \checkmark   \\
\hline
\checkmark & \checkmark  & $m_{\varphi_2}^2\sim \Lambda_{\text{gc}}^4\ll m_{\varphi_1}^2\,,\, m_{\varphi_3\,}^2\sim \Lambda_{\text{ws}}^4$  & $\times$ & $\times$ \\
\hline
\checkmark &  $\times$  & $m_{\varphi_1}^2\sim\Lambda_{\text{QCD}}^4\ll m_{\varphi_2}^2\sim \Lambda_{\text{gc}}^4\ll  m_{\varphi_3}^2\sim\Lambda_{\text{ws}}^4 $ & (\checkmark)& $\times$  \\
\hline
\end{tabular}
\caption{Three-axion scenarios summarizing presence of gaugino condensation (GC), mass hierarchies, strong CP resolution, and dominant gauge couplings.}
\label{tab:upshot}
\end{table}

In this work, we will extend these studies and focus on the heterotic string axiverse and its discovery potential. In particular, we are interested in axions that couple to hidden gauge fields via CS couplings and may thereby produce gravitational waves during inflation, realizing the spectator axion mechanism. To that end, we characterize the generic structure of the axion mass spectrum and the effective couplings of heterotic axions beyond the QCD axion candidate to Abelian and non-Abelian gauge fields and discuss their implications for both cosmology and particle phenomenology. We also discuss the impact of generating non-perturbative potentials for the the non-QCD axions onto the CP quality problem of the QCD axion candidate. 

\cref{tab:upshot} summarizes the upshot of our results: it combines the structure of the non-perturbative quantum effects providing the axion mass with the QCD instanton contribution and the constraints from maintaining CP quality as well as the rather strict upper bound on the compactification volume imposed by heterotic perturbativity. The resulting axion mass spectrum is quite different compared to the type IIB or M-theory axiverses -- most of the 2-form axions here stay rather heavy, while typically the axion responsible for solving the QCD CP problem (if possible) is the lightest axion state with a mass scale proportional to $\Lambda_{\text{QCD}}^4$. The only exception arises for fibred CY compactifications which break the hidden $E_8$ gauge group completely via gauge bundle choice and/or Wilson lines, and with their two dominant K\"ahler moduli stabilized in a highly anisotropic regime. For this rather special case, a suppressed world-sheet instanton direction can arise providing a single fuzzy-dark matter (FDM) candidate among the heterotic string axions.

The paper proceeds as follows: First, in~\cref{sec:heteroticreview}, we review revelant details about heterotic string theories and their compactifications. Then in~\cref{sec:hetaxions} we formulate the effective action for axions in heterotic compactifications and analyze their mass spectrum and couplings. Finally, in~\cref{sec:conclusions} we conclude.


\section{Review of the Heterotic String}
\label{sec:heteroticreview}

Of the five perturbative superstring theories, weakly coupled heterotic string theory is particularly appealing from a phenomenological perspective as it naturally accommodates key features of the Standard Model within a consistent high-energy framework. Its gauge sector arises from either a ten-dimensional $\left( \mathrm{E}_{8} \times \mathrm{E}_{8} \right) \rtimes \mathbb{Z}_{2}$ or  $\mathrm{Spin}(32)/\mathbb{Z}_2$\footnote{Sometimes refered to as $\mathrm{SemiSpin}(32)$~\cite{McInnes:1999va,McInnes:1999pt}.} symmetry, allowing for grand unified theories (GUTs) and the embedding of realistic gauge groups after compactification. Compactifications on Calabi-Yau threefolds with suitable vector bundles can yield chiral spectra, which are essential for reproducing the observed particle content. From its discovery~\cite{Gross:1984dd, Gross:1985rr}  there have been many works constructing a 4D low-energy EFT which matches the minimal supersymmetric standard model (MSSM)~\cite{Braun:2005ux,Buchmuller:2005jr, Lebedev:2006kn,Lebedev:2007hv,Anderson:2011ns}. 

The heterotic worldsheet conformal field theory is built by combining left-moving modes of the 26-dimensional bosonic string with right-moving modes of the 10-dimensional superstring, in such a way that the resulting theory is consistent in ten spacetime dimensions.
Concretely, the right-moving (antiholomorphic) sector describes ten-dimensional supersymmetric fields: spacetime bosons $X^\mu(\bar{z})$ and their superpartners $\psi^\mu(\bar{z})$ for $\mu = 0, \dots, 9$. The left-moving (holomorphic) sector, instead, includes only the bosonic coordinates $X^\mu(z)$, and to fill the mismatch in central charges and ensure conformal invariance, one introduces an internal set of 16 extra left-moving degrees of freedom, $\Xi^I(z)$, where $I = 1, \dots, 16$.  One can write them as real fermionic coordinates, corresponding to 32 real left-moving worldsheet fermions, or as 16 complex fermions. Regardless of this choice, these are the fields which generate the gauge sector excitations of the heterotic string.

Modular invariance restricts the allowed choices for the compactification lattice of these internal degrees of freedom. The only even self-dual lattices in 16 Euclidean dimensions are the $E_8 \times E_8$ and $\text{Spin}(32)/\mathbb{Z}_2$ lattices. These yield two consistent heterotic string theories, both living in $D=10$ dimensions: the $E_8 \times E_8$ heterotic string, and the $\mathrm{Spin}(32)/\mathbb{Z}_2$ heterotic string. Upon compactification to four dimensions, these internal gauge symmetries give rise to non-Abelian gauge groups and moduli. 

To preserve $\mathcal{N}=1$ supersymmetry in four dimensions, the compactification manifold is required to be a Calabi-Yau threefold: a compact K\"ahler manifold with vanishing first Chern class and SU(3) holonomy. This ensures the existence of a single covariantly constant spinor, which is necessary for a single unbroken 4D supersymmetry.
In such compactifications, the ten-dimensional spacetime decomposes as
\begin{equation}
    \mathbb{R}^{1,3} \times \text{CY}_3\,,
\end{equation}
and the gauge bundle is chosen to satisfy the Hermitian Yang-Mills equations. Supersymmetry and anomaly cancellation constrain this bundle: its field strength $F$ must obey $F_{(0,2)} = F_{(2,0)} = 0$ and $g^{i\bar{j}}F_{i\bar{j}} = 0$, known as the DUY equations (Donaldson-Uhlenbeck-Yau)~\cite{Donaldson:1985zz,Uhlenbeck:1986ntc}, and the Bianchi identity for the NS-NS three-form $H$ reads
\begin{equation}
\label{Bianchi}
    dH = \frac{\alpha'}{4} \left( \text{tr} \, R \wedge R - \text{tr} \, F \wedge F \right)\,,
\end{equation}
requiring a non-trivial relation between geometry and gauge flux.

The simplest examples of CYs arise as hypersurfaces in toric varieties. One common construction is that of smooth hypersurfaces $X$ in complex projective space $\mathbb{CP}^{d+1}$ defined by the vanishing of a homogeneous polynomial of degree $k = d + 2$. These are sections of the line bundle $\mathcal{O}_{\mathbb{P}^{d+1}}(k)$. For $X$ to be Calabi-Yau, it must have trivial canonical bundle, even if the ambient toric variety does not. This ensures the existence of a globally defined holomorphic $(d,0)$-form $\Omega$ which is equivalent to the demands of Ricci flatness and SU(3) holonomy. 

The canonical bundle $K$ is the line bundle of holomorphic top forms 
$$\Omega_U(z_1, ..., z_d) dz_1 \wedge \cdots \wedge dz_d$$
on a patch $U$. On a smooth variety, $K$ is trivial if and only if the first Chern class vanishes. For hypersurfaces, this can be verified using the adjunction formula. Consider $X$ as a hypersurface in an ambient space $A$ of dimension $d$. The tangent bundle $T_{A}|_X$ splits as:
\begin{equation}
0 \to T_X \to T_A|_X \to N_X \to 0,
\end{equation}
where $N_X \cong \mathcal{O}_A(X)|_X$ is the normal bundle. The Chern classes satisfy:
\begin{equation}
    c(T_X) = \frac{c(T_A)}{c(N_X)}.
\end{equation}
If the ambient space is $\mathbb{CP}^{d+1}$, each homogeneous coordinate $z_i$ corresponds to a divisor $H$ with line bundle $\mathcal{O}(1)$, so $c(T_A) = (1 + H)^{d+2}$. A degree-$k$ hypersurface corresponds to $\mathcal{O}(k)$, so:
\begin{align}
    c(T_X) &= \frac{(1 + H)^{d+2}}{1 + kH} = (1 + (d+2)H + \cdots)(1 - kH + k^2H^2 - \cdots).
\end{align}
We then find:
\begin{equation}
    c_1(T_X) = (d + 2 - k) H.
\end{equation}
Requiring $c_1 = 0$ gives $k = d + 2$, the Calabi-Yau condition.

\subsection{Gauged Linear Sigma Models (GLSMs)}

The low energy effective worldsheet theory of the heterotic string compactified on a CY three-fold $X$ is a nonlinear sigma model (NLSM) with $\mathcal{N}=(0,2)$ supersymmetry, describing maps from the string worldsheet into the target space  $M^{1,3} \times X$~\cite{Witten:1993yc}. The internal geometry and gauge bundle data appear in the NLSM through the metric $G$, the 2-form gauge field $B$ and couplings to worldsheet fermions. 
While phyiscally relevant, the NLSM is strongly coupled in the UV, which makes it complicated to deal with. Therefore, we use gauged linear sigma models (GLSMs) as UV completions, which are 2d supersymmetric gauge theories that flow in the IR to NLSMs. For compactifications preserving $\mathcal{N}=1$ supersymmetry in four dimensions, we need to look at $\mathcal{N}=(0,2)$ GLSMs on the worldsheet. The field content includes the following multiplets:
\begin{itemize}
    \item Chiral multiplets $\Phi^i = (\phi^i, \psi^i_-)$ containing a complex scalar and a right-moving fermion, 
    \item Vector multiplets $V$ for each gauged Abelian symmetry ($U(1)^m$), that contain gauge fields $A$, gauginos, and a complex scalar $\sigma$.
    \item Fermi multiplets $\Lambda^\alpha=(\lambda^\alpha, F^\alpha)$ with a left-moving fermion and an auxiliary field $F$.
    \item Twisted chiral multiplets that encode the gauge field strengths and couple to Fayet-Illiopoulos parameters and theta angles. 
\end{itemize}
The $\phi^i$ fields can be viewed as coordinates on $\mathbb{C}^N$. The scalar potential includes F- and D-terms, that impose moment map constraints that define a symplectic quotient $\mathbb{C}^N/U(1)^m$: 
\begin{equation}
U(\phi, \sigma) = \sum_I \frac{1}{2e_I^2} D_I^2 + \sum_i |F_i|^2 + 2 \sum_I |\sigma_I|^2 \sum_i Q_i^{I2} |\phi_i|^2,
\end{equation}
with D- and F-terms given by:
\begin{equation}
D^I = \sum_i Q_i^I |\phi_i|^2 - a^I, \quad F_i^* = \left. \frac{\partial W}{\partial \Phi_i} \right|_{\theta = 0}.
\end{equation}
The moduli space of the classical vacuum is then a toric variety, while adding a gauge-invariant transverse superpotential defines a hypersurface within the ambient space. 
To ensure that the resulting target space is a Calabi-Yau, the gauge charges $Q^i$ must sum to 0: $\sum_i Q^i=0$, guaranteeing that the first Chern class of the tangent bundle vanishes. This also enforces conformal invariance of the IR theory, in the absence of anomalies. 

In $\mathcal{N}=(0,2)$ theories, the Fermi multiplets are subject to chiratility constraints
\begin{equation}
   \bar{ \mathcal{D}}_+\Lambda^\alpha=\sum_i E_i^\alpha (\Phi)\Psi^i\,,
\end{equation}
where $E_i^\alpha(\Phi)$ are holomorphic functions of the chiral superfields. These functions encode the geometry of a holomorphic vector bundle $V$ over the CY, where the cohomology of $V$ is described by the surviving massless fermions (the ones that are not gauged away or become massive). 
These bundles usually are usually encoded in a so-called \textit{monad construction} expressed as an exact sequence of holomorphic vector bundles~\cite{Blumenhagen:2006ci, Anderson:2008uw}: 
\begin{equation}
0 \to \mathcal{O}^{\oplus p} \xrightarrow{M} \bigoplus_\alpha \mathcal{O}(Q^\alpha) \xrightarrow{N} \bigoplus_m \mathcal{O}(-q_m) \to 0\,.
\end{equation}
In the above equation, $\mathcal{O}^{\oplus p}$ denotes a trivial bundle of tank $p$, corresponding to uncharged Fermi multiplets, $\mathcal{O}(Q^\alpha)$ and $\mathcal{O}(-q^m)$ are direct sums of the line bundles over the CY characterized by integer charge vectors $Q^\alpha$ and $q^m$ under the GLSM gauge symmetries. The maps $M$ and $N$ encode the holomorphic data determined by the superpotential couplings and the chirality constraints. 

The vector bundle defined this way reads  as the cohomology of the complex:
\begin{equation}
    V=\frac{\ker(N)}{\im (M)}\,.
\end{equation}
When $p=0$ we have a \textit{minimal monad}, and $V=\ker(N)$, while $p>0$ allows for more general structures. 
This construction is used in heterotic compactifications to construct stable bundles. To ensure $c_1(V)=0 $ (consistent with supersymmetry) we need to impose the condition 
\begin{equation}
    \sum_\alpha Q^\alpha =\sum_m q^m\,,
\end{equation}
while to ensure that $c_1(TX)=0$ (CY condition), we require 
\begin{equation}
    \sum_i Q^i=0\,.
\end{equation}

\subsection{The gauge group}
For simplicity, we focus on the $E_8 \times E_8$ heterotic string as the the ten-dimensional gauge symmetry has the attractive feature of factorizing cleanly into two separate $E_8$ sectors. Upon compactification, realistic gauge groups can be engineered by appropriately embedding the internal gauge bundle into one of the $E_8$ factors. Specifically, one chooses a structure group $G_1 \subset E_8$ for the internal bundle $V_1$, and 
the unbroken gauge symmetry in four-dimensions is given by the commutant of $G_1$ in $E_8$
\begin{equation}
    E_8 \supset G_1 \times H_4
\end{equation}
where $H_4$ is the visible 4d gauge group. 
The choice of structure group thus determines the visible gauge group~\cite{Anderson:2011ns}, as well as the matter content, which is encoded in the decomposition of the adjoint representation of $E_8$ under $G_1\times H_4$, and the cohomology of the associated bundle-valued representation.

Typical embeddings used to get 4d gauge groups that resemble out universe are:

\begin{itemize}
    \item SU(5) GUT: \\
    Structure group: $G_1 = SU(5)$ \\
    Commutant: $H_4 = SU(5)$ \\
    This yields a grand unified theory with the usual $\mathbf{10} \oplus \overline{\mathbf{5}}$ matter content. Further breaking to the Standard Model can occur via Wilson lines.

    \item SO(10) GUT:  \\
    Structure group: $G_1 = SU(4)$ \\
    Commutant: $H_4 = SO(10)$ \\
    This setup has the advantage of allowing each SM generation to fit into a single $\mathbf{16}$ spinor of $SO(10)$.

    \item E$_6$ : \\
    Structure group: $G_1 = SU(3)$ \\
    Commutant: $H_4 = E_6$ \\
   Here, the visible gauge group is large, which requires additional breaking mechanisms.

    \item Pati-Salam: \\
    Structure group: $G_1 = SU(4) \times SU(2)_R$ \\
    Commutant: $H_4 = SU(2)_L \times SU(2)_R \times SU(4)$ \\
    These models unify quarks and leptons at an intermediate scale.

    \item Standard Model-like: \\
    Structure group: $G_1 = SU(3) \times SU(2) \times U(1)^n$ or more elaborate constructions. \\
    Commutant: $H_4 \supset SU(3)_c \times SU(2)_L \times U(1)_Y$ \\
    The low energy group resembles the SM group, though careful engineering (e.g., via fluxes or Wilson lines) is needed to obtain the correct spectrum and couplings.
\end{itemize}

The low-energy chiral spectrum is determined by decomposing the adjoint representation $\mathbf{248}$ of $E_8$ under $G_1 \times H_4$, and computing the cohomology associated with the resulting bundle-valued representations.

To realize these constructions, we take a Calabi-Yau threefold $X$ described by a complete intersection in a toric variety, with $h^{1,1}=k$ independent K\"ahler parameters. 
A line bundle $L$ on $X$ is completely specified by its first Chern class $c_1 \in H^{2}(X,\mathbb{Z})$, which can be expanded in a basis $\{\omega_i\}$ of $H^2(X,\mathbb{Z})$, as
\begin{equation}
    c_1(L)=\sum_{i}^{h^{1,1}}n_i\omega_i\,, \qquad L=\mathcal{O}(n_1,\dots,n_k)=\mathcal{O}\left(\sum_i n_i \omega_i\right)\,.
\end{equation}
where $n_i \in \mathbb{Z}$. 
More general vector bundles can be constructed as cohomologies of complexes, as reviewed above. A vector bundle $V$ of rank $r$ can be defined via 
\begin{equation}
    0\rightarrow \mathcal{O}^{\oplus p}\xrightarrow{M} \bigoplus_{\alpha=1}^{r_1}\mathcal{O}(n_i ^\alpha)\xrightarrow{N}\bigoplus_{m=1}^{r_2}\mathcal{O}(m_i^m)\rightarrow 0.
\end{equation}
This defines the vector bundle $V$ with rank ${\rm rk} (V)= r_1-r_2$, as $V = \frac{\text{Ker}(N)}{\text{Im}(M)}$ with $p \geq 0$. 
The total Chern class is given by:
\begin{equation}
    c(V) = \prod_{\alpha, m}\frac{1+\sum_i n_{i}^\alpha\omega_i}{1+\sum_i m_i^m\omega_i}.
\end{equation}
In particular, the first Chern class reads:
\begin{equation}
    c_1(V) = \sum_{i=1}^{h^{1,1}} \left(\sum_{\alpha=1}^{r_1} n_{i}^\alpha -\sum_{m=1}^{r_2} m_i^m \right)\omega_i.
\end{equation}
For an $SU(N)$ vector bundle, supersymmetry requires $c_1(V)=0$. Given this setup, one can compute the massless spectrum by evaluating the cohomology groups associated to the bundle-valued representations.

\paragraph{Bianchi identity} 
In addition, consistency of the string background imposes the Bianchi identity \cref{Bianchi}, which relates the geometry of the compactification to the topology of the gauge bundle.
Upon integration over compact 4-cycles in the Calabi–Yau, this identity translates into a topological condition involving second Chern classes~\cite{Green:1984sg}:
\begin{equation}
    \sum_ic_2(V_i)=c_2(TX)\,,
\end{equation}
for non-Abelian bundles. In models with Abelian bundles, the condition generalizes to:
\begin{equation}
    \sum_{i=1}^K \text{ch}_2(V_{n_i}) + \sum_{m=1}^M a_m c_1^2(L_m) = -c_2(T_X).
\end{equation}
The constants $a_m$ appearing in the Bianchi identity for Abelian bundles are not arbitrary:
they are group-theoretic coefficients that depend on how the Abelian $U(1)^m$ factors are embedded in the ten-dimensional gauge group. 
To determine them, we expand the internal gauge field in the Cartan subgroup of $E_8$, so that the embedding of each Abelian factor is specified by a charge operator  $Q_m\in \mathfrak{e}_8$, and fields carry charges $q_m \in \mathbb{Z}$ under this generator.
The normalization of the trace is defined through 
\begin{equation}
    \text{tr}(F_m^2) = a_m \, c_1^2(L_m),
\end{equation}
where $\text{tr}$ is the trace in the adjoint representation of $E_8$. The value of $a_m$ is given by:
\begin{equation}
    a_m = \frac{1}{4} \, \text{Tr}_{\text{adj}}(Q_m^2),
\end{equation}
where the trace sums the squared charges of the adjoint representation. 
Alternatively, we can relate $a_m$ to the level of embedding $k$ of the $U(1)$ into $E_8$ by expressing the normalization in terms of the Kac-Moody level, as 
\begin{equation}
    a_m=\frac{k_m}{30}\,,
\end{equation}
reflecting the standard trace identity in $E_8$: 
\begin{equation}
     {\rm tr}_{\mathbf{248}}(T^aT^b)=k \delta^{ab}\,.
\end{equation}
The Kac-Moody level $k_m$ counts how the $U(1)$ charges appear in the decomposition of the adjoint of $E_8$ and determines the normalization of kinetic terms and anomaly coefficients.

\paragraph{DUY equations}
In addition to the topological constraints from the Bianchi identity, the D-term equations, derived from supersymmetry, impose the so called DUY conditions, which require that each slope-stable vector or line bundle has vanishing slope
\begin{equation}
    \int_X J\wedge J\wedge c_1(V_{n_i}) = 0, \quad \int_X J\wedge J\wedge c_1(L_m) = 0,
\end{equation}
with one-loop corrections when $c_1(V) \neq 0$. These constraints further restrict the allowed moduli, freezing some combinations of the K\"ahler moduli and the dilaton. 
The axions dual to those directions become longitudinal components of massive $U(1) $ gauge bosons via the Green-Schwarz mechanism, and the effective theory retains only anomaly-free gauge symmetries.

\subsection{Heterotic EFT in 10D and 4D}

Axions in heterotic string theory come from the Kalb-Ramond 2-form $B_2$~\cite{Witten:1984dg}.
The relevant part of the 10D heterotic action is
\begin{equation}
\label{10d:L}
\begin{aligned}
    \mathcal{L}_{10D} &= \frac{1}{2\kappa_{10}^2} \sqrt{-g} R - \frac{1}{4\kappa_{10}^2} H \wedge \star H - \frac{\alpha'}{8\kappa_{10}^2} \text{tr}(F \wedge \star F) \\
    &= \frac{2\pi}{g_s^2 \ell_s^8} \sqrt{-g} R - \frac{2\pi}{g_s^2 \ell_s^4} \cdot \frac{1}{2} H \wedge \star H - \frac{1}{4(2\pi) g_s^2 \ell_s^6} F \wedge \star F\,,
\end{aligned}
\end{equation}
where $H = dB + \omega_{3L} - \omega_{3Y}$ and the trace refers to the adjoint of $E_8 \times E_8$ or $SO(32)$.

Compactifying to four dimensions and integrating over the internal manifold $X$ with physical volume ${\rm Vol}(X)=\mathcal{V}\ell_s^6$ yields the effective action
\begin{equation}
    S_{4D} \supset \frac{M_{\text{Pl}}^2}{2} \int d^4x \sqrt{-g} R - \frac{1}{4 g_{\text{YM}}^2} \int d^4x\, F \wedge \star F - \frac{2\pi \mathcal{V}}{g_s^2 \ell_s^4} \int \frac{1}{2} H \wedge \star H\,,
\end{equation}
with the four dimensional parameters $M_p^2=4\pi\frac{{\rm Vol}(X) }{g_s^2\ell_s^8}=4\pi\frac{\mathcal{V} }{g_s^2\ell_s^2}$ and $g_{YM}^2=4\pi \frac{g_s^2\ell_s^6}{{\rm Vol}(X)}=4\pi \frac{g_s^2}{\mathcal{V}}$.
Therefore, we can write $\alpha_{YM}=g_{YM}^2/4\pi$ as 
\begin{equation}
    \alpha_{YM}=\frac{g_s^2}{\mathcal{V}}\,.
\end{equation}
If we allow for a non-standard embedding of the SM into the heterotic string at Kac-Moody level $k>1$, then in general (see Witten for more) 
\begin{equation}
\label{alphagut}
    \alpha_{GUT}=\frac{\alpha_{YM}}{k}=\frac{g_s^2}{k \mathcal{V}}\,.
\end{equation}
The string scale $M_s=1/\ell_s$ can be evaluated to be $M_{s}=(k \alpha_G/4\pi)^{1/2} M_p$, where $\alpha_G$ is the strong coupling constant, such that if $\alpha_G\sim 1/25$, then $M_s\sim M_p\sqrt{k}/18$ which is the usual perturbative heterotic string scale. 

We emphasize here a direct consequence of the above relation between the 4D gauge coupling, the CY volume and the string coupling $g_s$. Namely, the restriction to perturbative heterotic string theory $g_s \lesssim 1$ (implying the absence of e.g. M5-branes of heterotic M-theory) in combination with phenomenological requirement $\alpha_{YM}\simeq 1/25$ of gauge coupling unification of the the MSSM gauge couplings into the $E_8$ GUT structure implies a stringent upper bound on the compactification volume~\cite{Hebecker:2004ce,Cicoli:2013rwa}
\begin{equation}
    {\cal V}\lesssim 20 - 30\quad.
\label{eq:volconstraint}
\end{equation}
As we will see below, this crucially limits $h^{1,1}$ if we demand that all 2-cycle volumes satisfy $v^i\gtrsim 1$ to ensure control over the worldsheet instanton expansion. In highly anisotropic fibred CY compactifications,~\cref{eq:volconstraint} still places constraints by bounding the largest curve volume as $v\lesssim 50$ (modulo the numerical values of the intersection numbers appearing in the volume form).

\subsection{Heterotic axions in 4D}
Next, we discuss the top-down axion content of the theory. The heterotic string generically contains both: one model-independent axion $a$, which is the 4D dual of $B_{\mu\nu}$ and universally present in all compactifications, and many model-dependent axions $b_i$, arising from the internal components of the $B$-field, with $i=1,\dots,h^{1,1}$.
The \textbf{model-independent axion} is defined via dualization as
\begin{equation}
    a = 2\pi \int_{CY} B_6\,, \quad \text{with} \quad dB_6 = \star dB_2\,,
\end{equation}
while model-dependent axions arise from expanding $B$ in a basis of harmonic 2-forms $\{\beta_i\}$:
\begin{equation}
\label{eq:B_harmonic}
    B = \frac{1}{2\pi} \sum_i b_i(x)\, \beta_i\,, \qquad \text{with} \quad \int_{\Sigma_j} \beta_i = \delta_{ij}\,.
\end{equation}
We now analyze the couplings of these axions to gauge fields through the modified Bianchi identity for $H$.
In four dimensions, we can enforce this identity by treating $a$ as a Lagrange multiplier:
\begin{equation}
    S \supset \int a \left( dH + \frac{1}{16\pi^2}(\text{tr}\, R \wedge R - \text{tr}\, F \wedge F) \right)\,.
\end{equation}
Integrating out $H$ yields an effective action for $a$:
\begin{equation}
    S(a) = \int d^4x \left[ -\frac{1}{2} f_a^2 (\partial a)^2 + \frac{a}{16\pi^2} (\text{tr}\, F \wedge F - \text{tr}\, R \wedge R) \right]\,,
\end{equation}
where the axion decay constant is
\begin{equation}
    f_a^2 = \frac{g_s^4 }{2\pi \mathcal{V}}\,.
\end{equation}

This reproduces the expected structure of an axion with a Chern-Simons coupling, where the coefficient is determined by the underlying string parameters and internal geometry.

The \textbf{model dependent axions} arise as the 0-form valued coefficients of the $B_2$ expansion in the basis of harmonic 2-forms $\{\beta_i\}$ as in \cref{eq:B_harmonic}.
The kinetic terms arise from dimensional reduction of the $H \wedge \star H$ term in the 10D action. Defining the K\"ahler metric:
\begin{equation}
    \gamma_{ij} = \int_{CY} \beta_i \wedge \star \beta_j\,,
\end{equation}
the 4D kinetic action becomes:
\begin{equation}
    S_{\text{kin}} = -\frac{1}{2\pi g_s^2 } \int d^4x\, \frac{1}{2} \gamma_{ij} \partial_\mu b_i \partial^\mu b_j\,.
\end{equation}

These axions acquire couplings to gauge fields via the 10D Green-Schwarz anomaly cancellation mechanism~\cite{Green:1984sg, Witten:1984dg}, as described in the following section.

\subsection{Anomalies, axions and Green-Schwarz}

In string theory, irreducible anomalies cancel due to group-theoretic identities, while factorizable (Abelian or mixed) anomalies cancel via a generalized Green-Schwarz mechanism that involves axionic couplings to Chern-Simons terms
\begin{equation}
    S_{GS}=\frac{1}{48(2\pi)^5\alpha'}\int B\wedge X_8\,,
\end{equation}
with 
\begin{equation}
    X_8=\frac{1}{24}{\rm tr} F^4-\frac{1}{7200}({\rm tr}F^2)^2-\frac{1}{240} ({\rm tr}F^2)({\rm tr}R^2)+\frac18 {\rm tr}R^4+\frac{1}{32}({\rm tr}R^2)^2\,.
\end{equation}
Using tadpole cancellation condition from the Bianchi Identity, separating the two gauge sectors, the action dimensionally reduced reads
\begin{equation}
\begin{aligned}
\label{Green-schwarz}
S_{G S} & =\frac{1}{64(2 \pi)^5 \alpha^{\prime}} \int B \wedge\left(\operatorname{tr} F_1^2\right)\left(\operatorname{tr}_1 \bar{F}^2-\frac{1}{2} \operatorname{tr} \bar{R}^2\right) \\
& -\frac{1}{768(2 \pi)^5 \alpha^{\prime}} \int B \wedge\left(\operatorname{tr} R^2\right)\left(\operatorname{tr} \bar{R}^2\right) \\
& +\frac{1}{48(2 \pi)^5 \alpha^{\prime}} \int B \wedge\left[\operatorname{tr}_1\left(F \bar{F}\right)\right]^2 \\
& +\frac{1}{32(2 \pi)^5 \alpha^{\prime}} \int B \wedge \operatorname{tr}_1\left(F \bar{F}\right)\left(\operatorname{tr}_1 \bar{F}^2-\frac{1}{2} \operatorname{tr} \bar{R}^2\right) + (1\leftrightarrow2)\,,
\end{aligned}
\end{equation}
 where the overlined quantities refer to the internal ones, while the others are the 4D ones.
We are interested in the first line of the Green-Schwarz action \cref{Green-schwarz}, which, after expanding the B-field in harmonic forms, gives us the Chern-Simons coupling for the model dependent axions~\cite{Dine:1987xk},
\begin{equation}
    -\frac{1}{2\pi^2 4!}\sum_i \int_X \beta_i\left[-\frac{\operatorname{tr}R\wedge R}{2 }+ 2\operatorname{tr}_1  F\wedge F-\operatorname{tr}_2 F\wedge F\right] \int b_i \frac{\rm tr_1 F\wedge F}{16\pi^2} +(1\leftrightarrow 2)\,.
\end{equation}
We make use of the Bianchi identity, \cref{Bianchi},whose integral over any compact 4-cycle in the internal manifold vanishes due to Stokes' theorem, assuming no boundaries or localized sources, to rewrite:
\begin{equation}
    -\sum_i \int_X\beta_i\wedge \frac{1}{16\pi^2}\left(\operatorname{tr}_1  F\wedge F-\frac12 \operatorname{tr} R\wedge R\right)\int b_i\left(\frac{\operatorname{tr}_1  F\wedge F}{16\pi^2}-\frac{\operatorname{tr}_2  F\wedge F}{16\pi^2}\right)\,.
\end{equation}
The effective 4D CS couplings can be written as:
\begin{equation}
    \mathcal{L}_{\text{CS}} = \sum_i \frac{n_i}{16\pi^2} b_i(x) \left( \text{tr}_1 F \wedge F - \text{tr}_2 F \wedge F \right) \,,
\end{equation}
where the coefficients $n_i$ depend on the internal geometry and background fluxes through:
\begin{equation}
    n_i = \int_{X} \beta_i \wedge  \frac{1}{16\pi^2} \left( \text{tr}_1 F \wedge F - \tfrac{1}{2} \text{tr}\, R \wedge R  \right)\,.
\end{equation}  
Diagonalizing the K\"ahler metric and canonically normalizing the axions as $b_i \to \vartheta_i = f_i b_i$ with
\begin{equation}
    f_i^2 = \frac{\gamma_i}{2\pi g_s^2 }\,,
\end{equation}
we arrive at the Chern-Simons couplings:
\begin{equation}
    \mathcal{L}_{\text{CS}}=\frac{1}{16\pi^2}\left(\frac{\vartheta_a}{f_a} + \sum_i\frac{n_i}{f_i} \vartheta_i\right){\rm tr_1}F\wedge F+\frac{1}{16\pi^2}\left(\frac{\vartheta_a}{f_a}-\sum_i\frac{n_i}{f_i} \vartheta_i\right){\rm tr_2} F\wedge F\,.
\end{equation}
Canonically normalizing the gauge field $F\to g_{YM}\times F$, we find
\begin{equation}
 \mathcal{L}_{\text{CS}} = \sum_i \frac{\lambda_i}{4f_i} \vartheta_i \left( \text{tr}_1 F \wedge F - \text{tr}_2 F \wedge F \right)
\,, \qquad \lambda_i = \frac{n_i k \operatorname{Re}[f]}{2\pi^2}\,,
\end{equation}
where $k$ is the current algebra level coming from the definition of the traces  ${\rm tr} F\wedge F=2k {\rm \mathbf{tr}}F\wedge F$.

At tree level, the gauge kinetic function determined from the kinetic terms of $F$ and the CS coupling with $a$ is~\cite{Kaplunovsky:1994fg} 
\begin{equation}
    f = \frac{\mathcal{V}}{4 \pi g_s^2 } + i\frac{a}{4\pi^2} \,.
\end{equation}
However, at one-loop, different choices of internal gauge bundles $V$  embedded into the first and second $E_8$ factor yield distinct threshold corrections from the model dependent axion CS couplings. 
Since we expanded the $B_2$ in a basis of harmonic two-forms $\{\beta_i\}$ of the Calabi-Yau manifold as in \cref{eq:B_harmonic}, we can do the same for the K\"ahler form:
\begin{equation}
    J = 2\pi \sum_{i=1}^{h^{1,1}} v_i \, \beta_i\,,
\end{equation}
then the $b_i$, as dimensionless axions with periodicity $b_i \sim b_i + 2\pi$, and the $v_i$ are the volumes of the associated two-cycles, can be combined in the complexified K\"ahler moduli $T_i$:
\begin{equation}
    T_i = v_i + i b_i\,.
\end{equation}
Then the one-loop corrected gauge kinetic function at large $T^i$ is (see e.g.~\cite{Dixon:1990pc,Gukov:2003cy})
\begin{equation}
    f = \frac{\mathcal{V}}{4\pi g_s^2 } + i\frac{a}{4\pi^2} \pm \frac{1}{4\pi^2}\sum_i T^i n_i, \,,
\end{equation}
where the $\pm$ depends if we're looking at the visible or hidden gauge sector.
Thus, the effective CS coupling reads:
\begin{equation}
\begin{aligned}
\label{final_CS}
   \lambda_{i, visible}&\equiv \lambda_{i,v}=  k \frac{8 n_i g_s^2}{(\mathcal{V}\pi+g_s^2 v_i n_i)}\,,\\
   \lambda_{i, hidden}&\equiv\lambda_{i,h}=  -k \frac{8 n_i g_s^2}{(\mathcal{V}\pi-g_s^2 v_i n_i)}\,.
\end{aligned}
\end{equation}
This way different couplings of the axions with the visible and with the dark sector can arise.

\subsection{Volume bound and axion multiplicity}


To reproduce phenomenologically viable values for the unified gauge coupling as given in \cref{alphagut}, the Calabi–Yau volume in string units must satisfy
\begin{equation}
	\label{eq:VolBound20}
    \mathcal{V}=\frac{1}{6}\kappa_{ijk}v^iv^jv^k\lesssim 20\,,
\end{equation}
where the $v^i$ denote the K\"ahler parameters measuring volume of the 2-cycles in the internal manifold, and $\kappa_{ijk}$ denote the triple intersection numbers. 
This volume bound can have different implications depending on the topology of the internal space. 

In isotropic compactifications, where all 2-cycle volumes are of similar size $v^i\sim v\lesssim 3$, the number of non-vanishing intersection numbers grows as $\sim \frac16 \mathcal{O}((h^{1,1})^3)$~\cite{Cicoli:2011it}. Approximating $\kappa_{ijk}\sim\mathcal{O}(1)$, the volume constraint implies a bound on the combination of $h^{1,1}$ and $v$
\begin{equation}
    \mathcal{V}\simeq\frac{1}{36}(h^{1,1})^3 v^3\lesssim 20\,.
\end{equation}
This scaling arises because the triple intersection number is a rank-three totally symmetric tensor, which can have at most $\binom{n+3-1}{3}$ independent number of components, where $n=h^{1,1}$ is the dimension of the vector space $n$ which the tensor is defined ($H^{1,1})$.

\begin{figure}
    \centering
    \includegraphics[scale=0.8]{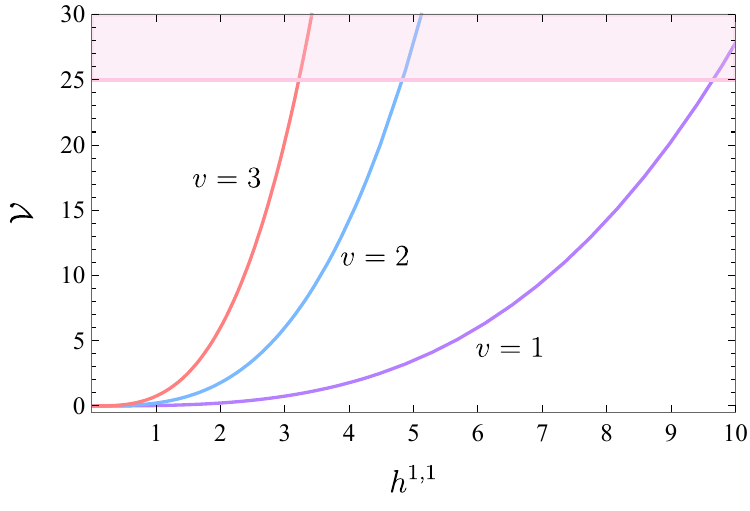}
    \caption{Volume bound for different values of the 2-cycle volume. The pink line shows the maximum allowed volume \cref{eq:VolBound20}.}
    \label{fig:volume_bound}
\end{figure}
Suppression of non-perturbative corrections, which is equivalent to control of the instanton series, requires $v^i\gtrsim \mathcal{O}(1)$. This puts an upper bound on the number of sizable 2-cycles which we show in \cref{fig:volume_bound}. 
There can be at most 8 axions with $v\sim1$, 4 axions if $v=2$, and so on. Extrapolating this would bound the number of axions to be less than 1 for $v\sim 8$, however when we have $\leq$ 2 axions the combinatorial property that gives us the $\frac16$ scaling no longer applies as it comes to compensate the identical permutations. 
We note that these are loose bounds, as intersection numbers can be as large as $\sim10$. Hence an isotropic heterotic compactification can
support at most a handful of axions.

One might hope to evade this restriction in a highly anisotropic compactification, such as a ``Swiss Cheese" Calabi-Yau with volume~\cite{Conlon:2005ki}
\begin{equation}
    \mathcal{V}=\tau_b^{3/2}-\sum_{s}\tau_s^{3/2}\,,
\end{equation}
where $ \tau_b $ denotes the volume of a large 4-cycle and the $ \tau_s $ are small blow-up cycles.
  Because the $\tau_s$ appear with negative signs, taking
$\tau_b\approx\tau_s$ can keep the volume small while
\emph{individually} sending both $\tau_b$ and $\tau_s$ to large values.  
However, it seems unnatural to allow for this strategy to be unconstrained. We can provide several arguments suggesting that 4-cycle volumes cannot be arbitrarly large even if they keep the overall volume fixed. First, consider a compactification with $h^{1,1}=2$, with $\tau_b=\kappa_{bbb}v_b^2$ and $\tau_s=\kappa_{sss}v_s^2$, such that the volume reads 
\begin{equation}
    \mathcal{V}=
    \left(\kappa_{bbb}\,v_b^{3}
          -\kappa_{sss}\,v_s^{3}\right)\,,
\end{equation}
Now impose the requirement
$\mathcal{V}\lesssim20$.  If one tries to take both
$\tau_b$ and $\tau_s$ parametrically large while keeping their difference
small, which forces
\begin{equation}
  v_b
  \;\simeq\;
\left(\kappa_{sss}/\kappa_{bbb}\right)^{1/3}\,v_s
  \;+\;
  \mathcal{O}\!\left(\frac13\mathcal{V}\,v_s^{-2}\right)\to \,r v_s\,,
  \label{eq:tbts}
\end{equation}
i.e.\ the two K\"ahler parameters must approach each other with scaling given by the ratio $r\equiv \left(\frac{\kappa_{sss}}{\kappa_{bbb}}\right)^{1/3}$. We now define a divisor class 
\begin{equation}
    L=D_b-\lambda D_s\,, \text{ with } \lambda>0\,.
\end{equation}
whose self intersection is $L^3= \kappa_{bbb}-\lambda \kappa_{sss}=0$ if $\lambda= \frac{1}{r}$. Geometrically, this means that $L$ lies on the boundary of the K\"ahler cone, as it is \textit{nef} but not \textit{ample}. In order words, $L\cdot \mathcal{C}\geq0$ for every effective curve $\mathcal{C}$ but it does not lie in the interior of the cone since there will be one intersection with an effective curve that vanishes. By duality of the nef and the Mori cone, any nef divisor on the boundary must have zero intersection with some effective curve class: there exists an effective curve $\mathcal{C}$ s.t. $L\cdot \mathcal{C}=0$. (Indeed $L^2$ is, intersecting two divisors which cuts out a holomorphic curve so basically $L^2$ is dual to $\mathcal{C}$). 

Now take an effective curve $\mathcal{C}$, with divisor intersections 
\begin{equation}
    a= D_b\cdot \mathcal{C}>0 , \qquad b= - D_s\cdot \mathcal{C}>0 \qquad \lambda \equiv\frac{a}{b}\,, 
\end{equation}
which will have the volume
\begin{equation}
    J\cdot \mathcal{C}= b v_s\left[ \lambda \frac{v_b}{v_s}-1 \right]\,.
\end{equation}
Therefore, in the limit \cref{eq:tbts}, the volume vanishes when $\lambda = v_s/v_b\to 1/r$. 

This is the curve whose existance is guaranteed by $L^3=0$, as it shrinks to zero volume in the large two-cycle limit. 
The above illustrates that one cannot send both 4-cycle volumes to infinity with fixed overall volume without hitting a boundary of the K\"ahler cone where an effective curve collapses. 

A similar argument can be made for more general swiss cheese structure manifolds.
Such a swiss cheese CY may have an intersection number structure such that we have $\tau_b=\kappa_{bij}v^iv^j= \kappa_{bbb}v_b^2+\kappa_{bbs}v_bv_s +\kappa_{bss}v_s^2$. In order to then rewrite the volume form in terms of $\tau_b$ and $\tau_s$ then one adds and subtracts an appropriate term $c v_s^2$ to complete the squares in the volume form, such that in terms of $\tau_b= (a v_b + b v_s)^2$ and $\tau_s= - c v_s^2$ the volume takes the swiss cheese form. 
The same result apply to this case, except that the linear relation between $v_b$ and $v_s$ gets modified 
\begin{equation}
 \mathcal{V}\sim (a v_b + b v_s)^3 - c v_s^3\sim \text{const}\,,\qquad  v_b =  v_s\frac{( c-b)}{a} +\mathcal{O}(\mathcal{V} v_s^{-2})\,.
\end{equation}

We now provide a simple explicit example illustrating this qualitatively.
Consider the blow-up of $\mathbb{P}^2$ at a point, with divisor basis $\{H, E\}$, where $H$ denotes the pullback of the hyperplane class and $E$ is the exceptional divisor: a $(-1)$ curve satisfying $E \cdot E = -1$. Let the K\"ahler form be parametrized as
\begin{equation}
J = v_H H - v_E E, \qquad v_H, v_E > 0,
\end{equation}
ensuring positivity of volumes for all effective curves. An important effective curve class on this surface is
\begin{equation}
\mathcal{C} = H - E,
\end{equation}
which corresponds to the proper transform of a line through the blown-up point. Although this class appears as a formal difference in the chosen basis, it is indeed an effective, rigid curve in the del Pezzo surface.

The volume of this curve is given by
\begin{equation}
\text{Vol}(\mathcal{C}) = J \cdot \mathcal{C} = (v_H H - v_E E) \cdot (H - E) = v_H - v_E.
\end{equation}
This shows that $\mathcal{C}$ becomes small as $v_H \to v_E$ from above, and shrinks completely at the boundary $v_H = v_E$ of the K\"ahler cone. While $\mathcal{C}$ is a genuine effective curve class, its volume depends on the \emph{difference} of two K\"ahler parameters.

Now take the case of multiple K\"ahler fields. 
Take $\tau_b=\kappa_{bjk} v^j v^k$, and $\tau_{s_i}=\kappa_{{s_i}jk}v^jv^k$, and take 
\begin{equation}
    \tau_b\sim \sum_i \tau_{s_i}\to \infty\,.
\end{equation}
Implying
\begin{equation}
    \kappa_{bjk}v^jv^k\sim \sum_i \kappa_{{s_i}jk}v^jv^k\quad.
\end{equation}
This  means that imposing the volume bound but keeping the four cycles big imposes one quadratic relation on the direction of growth of the 2-cycles: they must blow up proportionally with fixed ratios set by the intersection numbers.

The structure of the string loop corrections to the moduli K\"ahler potential provides another, and correlated, signal indicating a shrinking 2-cycle. Namely, sending 4-cycles to large volumes breaks down the EFT which comes from string loop corrections to the K\"ahler potential, $\delta K^{(g_s)}$, which behave as homogeneous functions of degree $-2$ in the 2-cycle volumes~\cite{Cicoli_2008}. This implies that if one sends a K\"ahler modulus $\tau \to \infty$ while keeping the overall volume $\mathcal{V}$ fixed, the corresponding 2-cycle volume $t$ also diverges, and hence $\delta K^{(g_s)} \sim v / \mathcal{V} \sim \tau^{1/2} / \mathcal{V}$ grows without bound. Although the scalar potential exhibits an \emph{extended no-scale structure} that ensures the cancellation of leading order contributions from such corrections when $\delta K^{(g_s)}$ is of degree $-2$, the subleading contribution $\delta V_2$ remains sensitive to their magnitude. Therefore, in this limit, the loop corrections to the scalar potential become large and the effective field theory breaks down.

\subsubsection{A note on geometry and bundles}
\label{note_geo_bun}

The structure of the gauge bundle in heterotic compactifications is intimately linked to the topology of the internal Calabi–Yau manifold, particularly its non-trivial cycles.
For vector bundle $V$ of rank $n$, the topological data is encoded in the chern classes $c_i(V)$. A holomorphic, stable in the sense of the slope, bundle, must satisfy the anomaly cancellation condition which related the second Chern class $c_2(V)$ to that of the CY $c_2(TX)$.
When talking about line bundles, we only need $ c_1(V) \in H^2(X)\sim $ divisors, counted by $h^{1,1}$. If $h^{1,1}$ is small, the options for embedding the gauge bundle are limited and you can't construct too complicated bundles. 

In many explicit constructions, especially those based on line bundles, the gauge bundle is written as a direct sum of line bundles over divisors:
$V=\bigoplus_{i}\mathcal{O}(D_i)$. Here, each divisor $ D_i $ corresponds to an element of $ H^2(X) $, whose dimension is counted by $ h^{1,1} $. Thus, the number of available divisors directly limits the flexibility in defining such bundles. When $ h^{1,1} $ is small, the space of line bundle configurations is highly constrained, making it difficult to construct bundles that satisfy anomaly cancellation and supersymmetry. Conversely, a larger $ h^{1,1} $ provides more geometric freedom to define richer bundle topologies. Therefore, while the rank of the bundle is not directly constrained by $ h^{1,1}$, constructing more intricate gauge bundles typically requires a compactification geometry with a greater number of independent cycles.\footnote{We thank Fabian Ruehle for explaining the content of \cref{note_geo_bun} to us and for reminding us the existence of shrinking curves in anisotropic limits of CY compactifications.} 


\section{Heterotic axion EFT: CP problem \& axion mass spectrum}
\label{sec:hetaxions}

\subsection{A short recap on heterotic moduli stabilization}

Discussion of an axion EFT in string theory requires addressing moduli stabilization within a given class of string compactifications. The strategies employed in type IIB string theory for compactification on warped conformal CY orientifolds with 3-form fluxes and 7-branes is of no use for the $E_8\times E_8$ perturbative heterotic string. Namely, we cannot avail ourselves of the presence of an RR-sector 3-form flux $F_3$ jointly with the NSNS 3-form flux $H_3$ in order to produce a flux discretuum by which we can fix both the c.s. moduli and the axio-dilaton and at the same time fine-tuning the resulting flux superpotential $W_0$ to be of small magnitude. 

Heterotic moduli stabilization in CY compactifications (including their orbifold limits in CY moduli space) has to proceed from the effective action determined by a K\"{a}hler potential and superpotential: 
\begin{equation}
    \begin{aligned}
    & K=-\ln(S+\bar S)-\ln{\cal V}(T_t,\bar T_i)-\ln\left(-i \int_{X}\Omega(z_a)\wedge \bar\Omega(\bar z_{\bar a})\right)+\Delta K_{\rm non-/pert.}\\
    & W= \int_{X} H_3\wedge\Omega(z_a) + \sum_i A_i e^{-a_i f_i(S,T_j,z_b)} +\sum_n B_k e^{-2\pi T_k}\quad.
    \end{aligned}
\end{equation}
Recalling the results of~\cite{deCarlos:1992kox,Gukov:2003cy,Gaillard:2007jr,Serone:2007sv,Anderson:2010mh,Dundee:2010sb,Parameswaran:2010ec,Cicoli:2013rwa,Leedom:2022zdm}, the first sum in $W$ parametrizes non-perturbative effects from gaugino condensation driven by unbroken non-Abelian gauge group factors surviving from the hidden $E_8$-factor. The second sum describes the contribution from worldsheet instantons. The contributions from gaugino condensation depend on the K\"ahler and c.s. moduli through the 1-loop threshold corrections which for largish values of the $T_i$ and $z_a$ depend linearly on those moduli, but are e.g. for simple toroidal orbifold limits of CYs dictated by modular invariance to appear in the form of the logarithm of the Dedekind eta function~\cite{Dixon:1990pc}. $\Delta K_{\rm non-/pert.}$ in turn represents perturbative $\alpha'$ and string loop as well as non-perturbative corrections to the K\"ahler potential~\cite{Shenker:1990uf,Candelas:1990rm,Barreiro:1997rp,Becker:2002nn,Anguelova:2010ed}.

In heterotic CY compactifications we can only use $H_3$-flux to fix the c.s. moduli. Its quantization produces either a VEV $W_0=|\langle \int H_3\wedge \Omega\rangle|\gtrsim {\cal O}(1)$ in the case of standard integer-quantized 3-form flux $H_3$, or at best $W_0=|\langle \int H_3^{fract.}\wedge \Omega\rangle|\gtrsim {\cal O}(0.1)$ in the case of $H_3$-flux due to the fractional CS term contribution e.g. from discrete Wilson lines ~\cite{Gukov:2003cy}.

In the absence of $H_3$-flux, generically a part of the c.s. moduli stabilization can happen at the SUSY Minkowski level ($D_zW=W=0$) by turning on a non-trivial  gauge bundle, already needed to break the visible $E_8$ towards the SM gauge group, as part of the background fields of the compactification. The resulting unbroken subgroup of $E_8\times E_8$ typically contains anomalous $U(1)$-factors whose D-terms will have the structure
\begin{equation}
    D=\frac{c}{S+\bar S}-q_C |C|^2
\end{equation}
where $c$ denotes the coefficient of the field-dependent FI term and $C$ are a summarily representation of the gauge bundle moduli which appear as SM gauge single chiral multiplets in the 4D EFT. Assume now that the dilaton $S$ is stabilized at a non-zero VEV with ${\rm Re}\,\langle S\rangle\simeq 2$ consistent with MSSM gauge coupling unification (more on this below). Then, the D-term scalar potential from the anomalous $U(1)$-s' D-terms will now drive some bundle moduli to acquire non-zero VEVs. It was shown that the combined moduli space of the c.s. moduli and bundle moduli has a partial cross structure~\cite{Anderson:2010mh,Anderson:2011cza,Anderson:2011ty} following an observation in~\cite{Witten:1985bz}. Hence, once the bundle moduli acquire non-zero VEVs from the D-terms, at least some of the c.s. moduli will be stabilized at zero VEV in turn. For certain CY manifold,s and in particular orbifolds, this mechanism can be sufficient to stabilize all of the c.s. moduli at the SUSY Minkowski level~\cite{Anderson:2010mh,Anderson:2011cza,Anderson:2011ty,Cicoli:2013rwa}. 

A variant of this situation arises if the 4D EFT of the given heterotic CY or orbifold compactification possesses a higher-order discrete $R$-symmetry $Z_N$ under which one the D-term chiral fields $C$ is charged. In this case, the superpotential may contain $R$-invariant high-order monomial terms $\Delta W\sim C^N$. As $C$ acquires a VEV of typical size $|\langle C\rangle|\sim \sqrt{c/{\rm Re}\,\langle S\rangle}\sim 0.1$, this induces an effective $W_0\sim \langle C\rangle^N \sim 10^{-N}$ which can be as small as ${\cal O}(10^{-10})$ for $Z_N$ with $N$ as large as 10~\cite{Dundee:2010sb}.

Next comes stabilizing the dilaton. Here, we can discriminate between two classes.
\begin{itemize}
    \item[\boxed{\text{GC}}] There is gaugino condensation~\cite{Veneziano:1982ah,Ferrara:1982qs} from unbroken non-Abelian gauge group factors surviving the breaking of the hidden $E_8$-factor~\cite{Font:1990nt,Nilles:1990jv}.
    \begin{itemize}
        \item[\textbullet] $W_0\simeq 0.1$ from fractional CS-invariants. The hidden $E_8$ now needs to remain unbroken, as only the Coxeter number of $E_8$ is large enough to stabilize the dilaton against $W_0$ in $D_SW=0$ at $\langle S\rangle\simeq 2$  required for gauge coupling unification~\cite{Gukov:2003cy,Cicoli:2013rwa}.
        \item[\textbullet] $W_0\ll 0.1$ from gauge bundle driven and D-term induced cs. moduli stabilization, producing high-order R-symmetry protected effectively constant terms in the superpotential. In this case, a lower-rank gauge group gaugino condensate surviving from the hidden $E_8$ can stabilize $S$ again near the phenomenologically desired value~\cite{Dundee:2010sb,Cicoli:2013rwa}.
        \item[\textbullet] $W_0=0$ after gauge bundle driven c.s. moduli stabilization. Dilaton stabilization via gaugino condensation now requires a racetrack (RT), i.e. two condensing non-Abelian gauge group factors surviving from breaking the hidden $E_8$~\cite{deCarlos:1992kox,Binetruy:1996uv}.
    \end{itemize}
    \item[\boxed{\text{noGC}}] The hidden $E_8$ gets completely broken by a combination of gauge bundle choice and additional Wilson lines to a surviving subgroup containing just several $U(1)$-factors. Gaugino condensation is now absent.
    \begin{itemize}
        \item[\textbullet] Dilaton stabilization has to proceed by a combination of perturbative and/or non-perturbative quantum corrections to the dilaton K\"ahler potential (such as the universally present `Shenker-like' terms~\cite{Shenker:1990uf}) which may produce a generically SUSY breaking $S$-minimum~\cite{Gaillard:2007jr,Leedom:2022zdm}.
    \end{itemize}
\end{itemize}
Finally, stabilization of K\"ahler moduli needs to proceed along similar lines classified by the presence or absence of gaugino condensation and a possible constant contribution $W_0$ to the superpotential.
\begin{itemize}
    \item[\boxed{\text{GC}}] Gaugino condensation occurs.
    \begin{itemize}
        \item[\textbullet] Stabilization of the K\"ahler moduli can proceed either via the dependence of the gaugino condensate on compactification moduli through threshold corrections to the gauge kinetic function~\cite{Dixon:1990pc}, or similar to the KKLT scenario, via worldsheet instanton corrections $\Delta W_{ws}\sim e^{-2\pi T_i}$ balancing against $W_0$ in the F-term condition $D_{T_i} \,W=0$.
        \item[\textbullet] Including the leading $\alpha'$-corrections to the volume moduli K\"ahler potential from 10D $R^2$ and $R^4$ curvature correction~\cite{Candelas:1990rm,Becker:2002nn,Anguelova:2010ed}, we can engineer an $\ell$VS-scenario like stabilization scheme for the K\"ahler moduli on CY manifolds whose CY volume takes the Swiss-Cheese form~\cite{Cicoli:2013rwa}. Here, the $\ell$ in $\ell$VS refers to the fact, that the total CY volume can at most be of ${\cal O}(20\dots 30)$ for perturbative ($g_s<1$) heterotic string compactification which maintain MSSM gauge coupling unification, so the CY volume can at best be 'large-ish' but not Large.
    \end{itemize}
    \item[\boxed{\text{noGC}}] No gaugino condensation.
    \begin{itemize}
        \item[\textbullet] K\"ahler moduli stabilization would now require at least one racetrack-like configuration of at two different worldsheet instantons for one volume modulus $T_i$ generating a minimum for it with non-vanishing $\Delta W_{\rm RT}(\langle T_i\rangle)$. This part of the superpotential can now act as an effectively constant $W_0$ against single worldsheet instantons for the remaining K\"ahler moduli to stabilize them similar to the KKLT scenario.
        \item[\textbullet] Alternatively, if $W_0\neq 0$ one can stabilize the K\"ahler moduli perturbatively given sufficiently many string loop and/or $\alpha'$-corrections to the volume moduli K\"ahler potential.
    \end{itemize}
\end{itemize}

\subsection{Sources of Axion Masses}
Axions in heterotic string compactifications generally acquire masses via three non-perturbative mechanisms:
\begin{enumerate}
    \item QCD instantons in the visible sector;
    \item Hidden sector gaugino condensation;
    \item Worldsheet instantons wrapping internal two-cycles.
\end{enumerate}
Assuming the visible sector is embedded in the first $E_8$, the QCD anomaly induces an axion potential:
\begin{equation}
    V_{\text{QCD}} = -\Lambda_{\text{QCD}}^4 \cos\left( \frac{\vartheta_a}{f_a} + \sum_i \frac{n_i}{f_i} \vartheta_i + \delta \right),
\end{equation}
where $\delta$ arises from the complex phase of the quark mass determinant.
We assume a superpotential of the form 
\begin{equation}
    W=W_0+W_{np}\,,
\end{equation}
where $W_0$ is a constant tree-level flux superpotential, and $W_{np}$ is the contribution coming from either worldsheet instantons or some condensing gauge group. 

If the second $E_8$ contains a non-Abelian subgroup $\mathcal{G}$ confining in the IR, gaugino condensation generates a superpotential in $\mathcal{N}=1$ SUGRA:
\begin{equation}
    W_{np} \sim \mathcal{A} e^{-\frac{8\pi^2}{c(\mathcal{G})}f_{\mathcal{G}}}
\end{equation}
where $f_{\mathcal{G}}$ and $c(\mathcal{G})$ are the gauge kinetic function and the dual Coxeter number of $\mathcal{G}$, respectively. In heterotic compactifications, one typically has $\text{Re}(f) \sim \mathcal{V}/(g_s^2 )$. For $\mathcal{G} = SU(N)$, $c(\mathcal{G}) = N$, 
This leads to the potential:
\begin{equation}
    V_{\text{gc}} = -\Lambda_{\text{gc}}^4 \cos\left( \frac{\vartheta_a}{f_a} - \sum_i \frac{n_i}{f_i} \vartheta_i \right),
\end{equation}
with
\begin{equation}
    \Lambda_{\text{gc}}^4 = \mu^4 \exp\left(-\frac{2\pi}{N} \frac{\mathcal{V}}{g_s^2}\right)\,.
\end{equation}
Worldsheet instantons wrapping holomorphic two-cycles generate non-perturbative contributions to the superpotential of the form
\begin{equation}
    W_{\text{np}} = \mathcal{A} \, e^{-2\pi T}
\end{equation}
where $T = v + i b$ is the complexified K\"ahler modulus, with $v$ the volume (in string units) of the wrapped cycle and $b$ its associated axion. These contributions induce a scalar potential for axions of the general form
\begin{equation}
    V_{\text{ws}} = -\Lambda^4 \cos\left( \sum_i \frac{c_i \vartheta_i}{f_i} \right), \qquad \Lambda^4 \sim \mu^4 \, e^{-2\pi v},
\end{equation}
where the $c_i$ are coefficients that depend on the specific instanton and its coupling to the axions. In the later sections we will restrict to the simplified case where each instanton only contributes to lifting one model dependent axion. 
The scale $\mu^4$ depends on the compactification and moduli stabilization data. In supergravity, the potential typically includes cross-terms of the form $V \sim e^K W_0 \mathcal{A} e^{-T}$, so $\mu^4$ often scales as $W_0 \mathcal{A}$. The flux superpotential $W_0$ typically lies between $10^{-13}$ and $10^{-1} M_{\text{Pl}}^3$, depending on the compactification, tuning, and origin of $W_0$.

We now explore how these contributions determine the mass spectrum in two- and three-axion systems, and how they affect the coupling structure. We begin with a general framework for understanding the hierarchy and role of each contribution:
\begin{enumerate}
    \item The QCD potential always contributes and generates a mass for the axion combination coupled to $\operatorname{tr}_1 F \wedge F$.
    \item Gaugino condensation contributes when non-Abelian hidden sector gauge groups are present. If the hidden $E_8$ is broken entirely to $U(1)$ factors, this contribution is absent.
    \item The internal Calabi-Yau volume $\mathcal{V}$ controls both $\alpha'$ corrections and gauge couplings. Realistic models require moderately large $\mathcal{V}$ (e.g., $\mathcal{V} \sim \mathcal{O}(10-20)$), limiting the emergence of ultra-light axions.
\end{enumerate}

\subsection{Strong CP problem}
To address the Strong CP problem arising from the CP-violating term in the QCD Lagrangian,
\begin{equation}
    \mathcal{L}_{\theta} = \frac{\theta_{\text{eff}}}{32\pi^2} \, \text{tr}(G \wedge G)\,,
\end{equation}
we invoke the Peccei–Quinn mechanism through a combination of axions that couple to the first $E_8$, which contains the visible-sector QCD gauge group. This specific linear combination of axions enters the QCD Chern--Simons term and therefore receives a potential from QCD instantons. The resulting potential dynamically minimizes the effective angle $\theta_{\text{eff}}$, driving it to zero. In this way, the axion field adjusts to cancel the CP-violating phase, providing a dynamical solution to the Strong CP problem.

It is important to emphasize, however, that the axion direction lifted by QCD instantons generally overlaps with those lifted by other non-perturbative effects, such as gaugino condensation or worldsheet instantons. These directions are not, in general, orthogonal in axion field space. To illustrate this, consider the following scalar potential:
\begin{equation}
\begin{aligned}
    V =& -\Lambda_{\text{QCD}}^4 \cos\left(\frac{\vartheta_a}{f_a} + \sum_i n_i \frac{\vartheta_i}{f_i} + \delta\right)
    - \Lambda_{\text{gc}}^4 \cos\left(\frac{\vartheta_a}{f_a} - \sum_i n_i \frac{\vartheta_i}{f_i}\right)
     \\
    & - \sum_{j=1}^{h^{1,1}} \Lambda_j^4 \cos\left(\sum_i c_i^{(j)} \frac{\vartheta_i}{f_i}\right)\,,
    \label{eq:V}
\end{aligned}
\end{equation}
where the scales $\Lambda_{\text{QCD}}, \Lambda_{\text{gc}}, \Lambda_j$ are defined in the previous subsection, and the $\vartheta_i$ denote model-dependent axions with decay constants $f_i$. 
In general, additional contributions from higher-order instanton effects, such as multi-instanton corrections, may also be present. These are typically suppressed by double exponentials and are therefore subleading compared to the single-instanton terms shown above. For the purposes of this analysis, we will neglect such higher-order corrections.

The theory described by \cref{eq:V} contains $N = h^{1,1} + 1$ axions and $N + 1$ leading terms in the potential.
In the regime where $\Lambda_{\text{QCD}}$ is the smallest scale, all axion vacuum expectation values are already fixed by the larger contributions from gaugino condensation and worldsheet instantons. As a result, the QCD-induced term is no longer able to dynamically relax $\theta_{\text{eff}}$ to zero, and the Peccei–Quinn mechanism fails to solve the Strong CP problem.

Let us consider an isotropic compactification, where all worldsheet instanton contributions are of comparable magnitude, and the non-perturbative scales exhibit the hierarchy
\begin{equation}
    \Lambda_{\text{ws}}^4 \gg \Lambda_{\text{gc}}^4 \gg \Lambda_{\text{QCD}}^4\,.
\end{equation}
In this setup, the worldsheet instanton potential $V_{\text{ws}}$ generically lifts $N = h^{1,1}$ axion directions. The remaining axionic degree of freedom is then fixed by the gaugino condensation term. As a result, by the time the QCD contribution becomes relevant, all axion vacuum expectation values are already stabilized, leaving no freedom to dynamically minimize the effective angle $\theta_{\text{eff}} = \vartheta_a + \sum_i n_i \vartheta_i + \delta$. In this case, the Strong CP problem is not solved dynamically, and the cancellation of $\theta_{\text{eff}}$ would require a fine-tuning of the axion vevs, which is no better than tuning the original $\theta$ angle itself.
The resolution lies in \textit{freeing} one axion vev so that it remains unfixed until the QCD contribution becomes dominant, allowing it to adjust and cancel the effective $\theta$-angle. To achieve this, we must ensure that only $N = h^{1,1}$ axion directions are lifted by effects stronger than QCD, while one direction remains light enough to be fixed by the QCD potential. This requires at least one non-perturbative contribution, either from gaugino condensation or a worldsheet instanton, to be more suppressed than $\Lambda_{\text{QCD}}$.

Before turning to explicit scenarios, let us comment on the constraints imposed by phenomenology. If the additional contribution is present but only slightly lighter than QCD, it can still interfere with the axion dynamics and shift the vev away from the CP-conserving minimum. 
To ensure that the Strong CP problem is reliably solved, the QCD contribution must dominate over any other source of explicit shift symmetry breaking for the axion. In particular, any subleading potential term must be suppressed relative to the QCD term by at least ten orders of magnitude, so that the induced shift in remains below current experimental bounds on the neutron electric dipole moment~\cite{Holman:1992us,Svrcek:2006yi,Kim:2008hd,Demirtas:2021gsq}: $\theta_{eff}\simeq\theta_{QCD}+\Delta\theta<10^{-10}$.   

Let us now consider the case in which a hidden-sector non-Abelian gauge group undergoes gaugino condensation at a scale below $\Lambda_{\text{QCD}}$. The associated contribution to the scalar potential takes the form
\begin{equation}
    \Lambda_{\text{gc}}^4 \sim W_0 M_s^3 \, e^{- \frac{2\pi}{N} \frac{\mathcal{V}}{g_s^2}} \ll 10^{-10} \times \Lambda_{\text{QCD}}^4 \sim 10^{-85} M_{\text{pl}}^4\,,
\end{equation}
where $W_0$ is the flux superpotential, $\mathcal{V} = {\rm Vol}(X) / \ell_s^6$ is the dimensionless Calabi--Yau volume in string units, and we take $\Lambda_{\text{QCD}}^4 \sim 10^{-75} M_{\text{pl}}^4$. Relating the string scale to the Planck scale via $M_s \sim g_s \mathcal{V}^{-1/2} M_{\text{pl}}$, we can express the gaugino condensation scale entirely in Planck units as
\begin{equation}
    \Lambda_{\text{gc}}^4 \sim W_0 \, \frac{g_s^3}{\mathcal{V}^{3/2}} \, e^{- \frac{2\pi}{N} \frac{\mathcal{V}}{g_s^2}}\,.
\end{equation}
Demanding $\Lambda_{\text{gc}}^4 \ll 10^{-85}$ imposes a stringent bound on the volume. Even taking optimistic values to minimize the contribution, such as $W_0 \sim 10^{-13}$, $g_s \sim 1$, and a minimal confining group with $N = 2$, one finds
\begin{equation}
    \Lambda_{\text{gc}}^4 \gtrsim 10^{-48} \gg 10^{-10} \Lambda_{\text{QCD}}^4\,,
\end{equation}
showing that gaugino condensation occurs at a scale vastly exceeding the QCD scale. As a result, any axion combination involved in this term will be stabilized well before the QCD potential becomes relevant. Thus, the axion vev is no longer free to adjust in response to the QCD contribution, and the Strong CP problem remains unsolved.
The only viable resolution in this case is to ensure that the hidden-sector gauge group does not confine. This can be achieved by breaking it to its Cartan subgroup, leaving only Abelian $U(1)$ factors, which do not undergo gaugino condensation and hence do not generate non-perturbative axion potentials.

Let us now consider the case here gaugino condensation occurs at a scale above QCD but the nearly free axion direction arises from a sufficiently low-scale worldsheet instanton contributing to the axion potential. Performing an analysis analogous to the gaugino condensation case, we find that the contribution takes the form
\begin{equation}
    \Lambda_{\text{ws}}^4 \sim W_0 \, \frac{g_s^3}{\mathcal{V}^{2/3}} \, e^{-2\pi v} \ll 10^{-85}
    \qquad \iff \qquad v \gtrsim 25\,,
\end{equation}
where $v$ denotes the volume of the wrapped two-cycle in string units. Achieving such a large suppression requires $v \gtrsim 25$, which is only possible in highly anisotropic compactifications, specifically, when one two-cycle is significantly larger than the others and dominates the total volume. This situation can naturally arise in fibred Calabi–Yau compactifications, where the base cycle is large and the fibre cycles remain small (of order unity).
\begin{figure}
    \centering
    \includegraphics[scale=0.8]{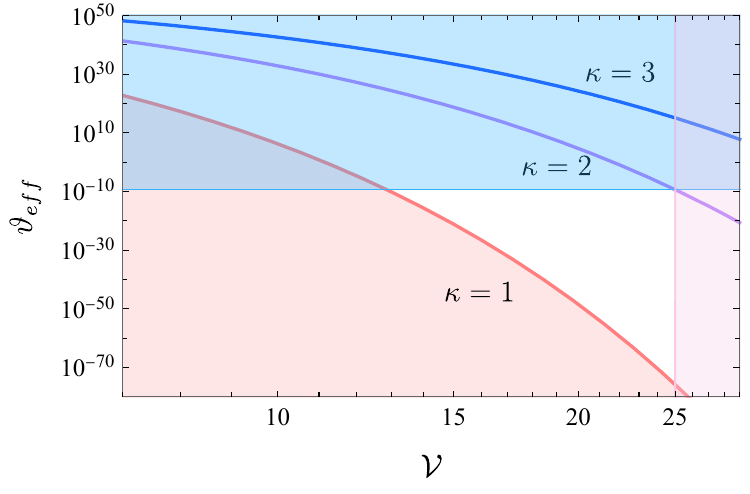}
    \caption{Effective $\theta$ angle arising from the inclusion of worldsheet instanton contributions, plotted as a function of the overall volume.  The shaded regions correspond to parameter values that are excluded. The red curve corresponds to $\kappa = 1$, with the associated exclusion region determined by the bound in \cref{eq:kappa}; the purple and blue curves represent $\kappa = 2$ and $\kappa = 3$, respectively. Higher values of the triple intersection number shift the curves upward, entering regions already excluded by observational and consistency constraints. The pink vertical line marks the upper bound on the volume, $\mathcal{V} < 25$, while the light blue horizontal line corresponds to the observational upper limit on the effective $\theta$ angle.}
    \label{fig:thetavsV}
\end{figure}
We can translate this in a bound on the volume, and so the total volume $\mathcal{V}$ is therefore bounded from both above and below. Since the 2-cycle volume $v$ scales as
\begin{equation}
    v = 2 \frac{\mathcal{V}}{\kappa} < \frac{50}{\kappa}\,,
\end{equation}
we obtain the following constraint on the total volume:
\begin{equation}
\label{eq:kappa}
    \frac{25\kappa}{2} \leq \mathcal{V} \leq 25\,.
\end{equation}
In \cref{fig:thetavsV}, we plot this angle as a function of $\mathcal{V}$ for different values of the triple intersection number $\kappa$. Compactifications with $\kappa > 2$ are already excluded by this analysis, as they would generate a $\theta_{\text{eff}}$ exceeding observational bounds. While the plot explicitly shows the allowed region for $\kappa = 1$, we omit the forbidden region for $\kappa = 2$ to avoid overshadowing the rest of the figure: in this case, the viable parameter space reduces essentially to a single point.

In the following subsections we will analyze the system in the simplest cases, when $h^{1,1}=1$ and $h^{1,1}=2$, where there are respectively, two or three axions. The first case, the simplest, can be analytically solved, and we can find the CS couplings of the physical axions, whereas the three-axion case is more complicated and cannot be solved fully analytically. 
We then provide examples for both cases, specifying when the Strong CP problem can be solved, when there can by a fuzzy dark matter axion candidate, and when the couplings between the hidden and the visible sector can be made different. 
We note that the examples we give are not complete models, as this would require more model building and case by case anlysis. 

\subsection{Two-Axion System}

We now consider a system of two axions: a model-independent axion $\vartheta_a$ and a model-dependent axion $\vartheta_1$, with decay constants $f_a$ and $f_1$, respectively.

Their kinetic terms and Chern-Simons couplings take the form:
\begin{align}
    \mathcal{L}_{\text{kin}} &= \frac{1}{2} (\partial \vartheta_a)^2 + \frac{1}{2} (\partial \vartheta_1)^2, \\
    \mathcal{L}_{\text{CS}} &= \frac{1}{16\pi^2} \left( \frac{\vartheta_a}{f_a} + \frac{n_1}{f_1} \vartheta_1 \right) {\rm tr}_1 F \wedge F
    + \frac{1}{16\pi^2} \left( \frac{\vartheta_a}{f_a} - \frac{n_1}{f_1} \vartheta_1 \right) {\rm{tr}}_2 F \wedge F.
\end{align}

We consider two subcases: one where we only consider QCD and gaugino condensation contributing to the potential and one where we only consider QCD and worlsheet instantons. 
This is because we have two axions, and in order to align their vevs to solve the Strong CP problem we cannot have more than two contributions.

\paragraph{\boxed{\text{GC}} }

We first consider the case where gaugino condensation is present, and the worldsheet instanton contribution is so small it can be safely neglected. We note that this case is merely a toy model, as for $h^{1,1}=1$ we are by definition in an isotropic compactification, where all two-cycles have similar volume, and thus there cannot be one single large two-cycle, effectively suppressing the worldsheet instanton contribution. 
We also note that in the realistic setup, where the worldsheet instanton case cannot be neglected, and gaugino condensation happens, the strong CP problem cannot be solved.
However, we still analyze this as it is an instructive case.  
This setup allows for a rotated field basis that diagonalizes the couplings:
\begin{align}
   \label{alpha}
   \varphi_1 &= \frac{1}{2}(\vartheta_a + \alpha \vartheta_1), \\
    \varphi_2 &= \frac{1}{2}(\vartheta_a - \alpha \vartheta_1),
\end{align}
with the field space `squashing parameter’
\begin{equation}
    \alpha = n_1 \frac{f_a}{f_1}.
\end{equation}
In this rotated basis, the CS couplings become:
\begin{equation}
    \mathcal{L}_{\text{CS}} = \frac{1}{8\pi^2 f_a} \left( \varphi_1 {\rm tr}_1 F \wedge F + \varphi_2 {\rm tr}_2 F \wedge F \right).
\end{equation}
The kinetic terms contain cross-terms proportional to $\alpha$:
\begin{equation}
    \mathcal{L}_{\text{kin}} = \frac{1}{2}\left(1 + \frac{1}{\alpha^2}\right)[(\partial \varphi_1)^2 + (\partial \varphi_2)^2] + \left(1 - \frac{1}{\alpha^2}\right) \partial \varphi_1 \partial \varphi_2.
\end{equation}
For $\alpha \approx 1$, these can be approximately diagonalized and normalized by a rescaling $\varphi_i \to \varphi_i \sqrt{\frac{1 + \alpha^2}{\alpha^2}}$.

In this limit, the visible and hidden sector couplings are aligned with orthogonal axion directions: $\varphi_1$ couples to the visible sector (QCD), and $\varphi_2$ to the hidden sector. The decay constants can be estimated as
\begin{equation}
    f_a \sim \frac{1}{\sqrt{\mathcal{V}}}, \qquad f_1 \sim \frac{1}{v},
\end{equation}
and for $n_1 \sim 2$, we find $\alpha \sim 1$, rendering the basis approximately orthonormal.
In the case where $\alpha\neq 1$, we need to rotate back to the original basis, but shift the decay constant of the model dependent axion in order to keep the axion periodicity:  
\begin{equation}
\qquad \tilde{f_1}= \frac{f_1}{\alpha}.
\end{equation}
The Chern-Simons couplings in this basis then read as 
\begin{equation}
\begin{aligned}
    \lambda_{\varphi_1,v}&= k\frac{8n_i g_s^2}{\mathcal{V}\pi-g_s^2 v_i n_i}\,,\\
    \lambda_{\varphi_2,h}&=-k \frac{8n_i g_s^2}{\mathcal{V}\pi - g_s^2 v_i n_i}\,.
\end{aligned}
\end{equation}

\paragraph{\boxed{\text{noGC}}} We now consider the case in which the hidden sector is broken to Abelian gauge groups only, so that no gaugino condensation occurs. In this case, axion masses arise solely from QCD instantons and worldsheet instantons. The relevant axion potentials are:
\begin{align}
    V_{\text{QCD}} &= -\Lambda_{\text{QCD}}^4 \cos\left( \frac{\vartheta_a}{f_a} + \frac{n_1}{f_1} \vartheta_1 + \delta \right), \\
    V_{\text{ws}} &= -\Lambda_{\text{ws}}^4 \cos\left( \frac{\vartheta_1}{f_1} \right).
\end{align}
The QCD term breaks the shift symmetry along the direction
\begin{equation}
    \varphi_{\text{QCD}} \propto f_1 \vartheta_a + n_1 f_a \vartheta_1,
\end{equation}
which we normalize to define an orthonormal basis:
\begin{align}
\label{phibasis}
    \begin{pmatrix} \varphi_1 \\ \varphi_2 \end{pmatrix} = U \cdot \begin{pmatrix} \vartheta_a \\ \vartheta_1 \end{pmatrix}, \quad U = \frac{1}{\mathcal{F}} \begin{pmatrix} f_1 & n_1 f_a \\ -n_1 f_a & f_1 \end{pmatrix}, \quad \mathcal{F} = \sqrt{f_1^2 + n_1^2 f_a^2}.
\end{align}
Here, $\varphi_1$ is the QCD axion, while $\varphi_2$ is orthogonal and receives a dominant mass from worldsheet instantons. Expanding the potential to quadratic order, the mass eigenvalues are approximately:
\begin{equation}
    m_{\varphi_1}^2 \simeq \frac{\Lambda_{\text{QCD}}^4}{f_a^2}, \qquad m_{\varphi_2}^2 \simeq \frac{n_1^2 \Lambda_{\text{QCD}}^4}{f_1^2} + \frac{\Lambda_{\text{ws}}^4}{f_1^2}.
\end{equation}
The anomaly coefficients and thus the CS couplings in the new basis become:
\begin{align}
    \frac{\vartheta_a}{f_a} + \frac{n_1 \vartheta_1}{f_1} &= \frac{\mathcal{F}}{f_a f_1} \varphi_1, \\
    \frac{\vartheta_a}{f_a} - \frac{n_1 \vartheta_1}{f_1} &= \frac{f_1^2 - n_1^2 f_a^2}{f_a f_1 \mathcal{F}} \varphi_1 - \frac{2 n_1}{\mathcal{F}} \varphi_2.
\end{align}
This implies:
\begin{equation}
    \mathcal{L}_{\text{CS}} = \frac{1}{16\pi^2} \left(\lambda_{\varphi_1,v} \frac{\varphi_1}{f_{\varphi_1}} \right) {\rm tr}_1 F \wedge F + \frac{1}{16\pi^2} \left( \lambda_{\varphi_1,h} \frac{\varphi_1}{f_{\varphi_1}} + \lambda_{\varphi_2,h} \frac{\varphi_2}{f_{\varphi_2}} \right) {\rm tr}_2 F \wedge F,
\end{equation}
where the effective decay constants are defined by demanding that the total cosine argument of the worldsheet instanton potential is periodic under simultaneous shifts of $\varphi_1$ and $\varphi_2$ \footnote{The worldsheet contribution is $\sim \cos(A\frac{\varphi_1}{f_{\varphi_1}}+B \varphi_2)$ where $A=f_a^2 n_1/\mathcal{F}^2$ and $B=\mathcal{F}^{-1}$, such that $f_{\varphi_2}=A/B$}
\begin{align}
    f_{\varphi_1} &= \frac{f_a f_1}{\mathcal{F}}, \\
    f_{\varphi_2} &= \frac{n_1 f_a^2 }{\mathcal{F}}.
\end{align}
After canonical normalization, the CS couplings become:
\begin{align}
\label{two_axions_CS}
    \lambda_{\varphi_1,v} &= k \cdot \frac{8 n_1 g_s^2}{\mathcal{V} \pi + g_s^2 v n_1}, \\
    \lambda_{\varphi_1,h} &=  -  k \cdot \frac{8 n_1 g_s^2}{\mathcal{V} \pi - g_s^2 v n_1} \cdot \frac{f_1^2 - n_1^2 f_a^2}{f_1^2 + n_1^2 f_a^2}, \\
    \lambda_{\varphi_2,h} &= + k  \cdot \frac{16 g_s^2 n_1^2 f_a^2 }{(\mathcal{V} \pi - g_s^2 v n_1)(n_1^2 f_a^2 + f_1^2)}.
\end{align}
These expressions determine how axions couple to the gauge sectors once gaugino condensation is absent and worldsheet effects dominate the hidden sector axion mass.
\begin{table}[t!]
\centering
\renewcommand{\arraystretch}{1.4}
\begin{tabular}{|c|c|c|c|c|}
\hline
\textbf{GC}& \textbf{Masses} & \textbf{CP} & \textbf{CS}   \\
\hline
$\times$  & $m_{\varphi_1}^2\sim \Lambda_{\text{QCD}}^4  \ll m_{\varphi_2}^2\sim\Lambda_{\text{ws}}^4$ & \checkmark  & $\lambda_{\varphi_11,v/h},\lambda_{\varphi_2,h}$ \\
\hline
\checkmark & $m_{\varphi_2}^2\sim \Lambda_{\text{gc}}^4\ll m_{\varphi_1}^2\sim \Lambda_{\text{ws}}^4$  & $\times$  & $\lambda_{\varphi_1,v} ,\, \lambda_{\varphi_2,h}$\\
\hline
\end{tabular}
\caption{Two-axion scenarios summarizing presence of gaugino condensation (GC), mass hierarchies, and Strong CP resolution, and dominant gauge couplings.}
\label{tab:table2}
\end{table}
The details of the two-axion system are summarized in \cref{tab:table2}.

\subsubsection{Illustrative example}

Consider now the setup introduced in~\cite{Svrcek:2006yi} where the CY is $X= C\times Y$ where $C$ is a Riemann surface with volume $\mathcal{V}_C$ and $Y$ a four-manifold with volume $\mathcal{V}_Y$. 
The integral therefore reduces to 
\begin{align}
    &- \frac{1}{16\pi^2} \int_C\beta \int_Y\left({\rm tr}_1 F\wedge F-\frac12 {\rm tr} R\wedge R\right)\int_M b_C\frac{{\rm tr}_1 F\wedge F}{16\pi^2}\\
    & -\frac{1}{16\pi^2} \int_Y\left({\rm tr_1} F\wedge F-\frac12 {\rm tr} R\wedge R\right)\int_M b_C\frac{{\rm tr_1} F\wedge F}{16\pi^2}\,.
\end{align}
where the last equality we used $\int_C \beta=1$. Therefore, we need to evaluate the integer 
\begin{equation}
\label{kprime}
    n=\frac{1}{16\pi^2}\int_Y \left({\rm\tr}_1 F\wedge F-\frac{1}{2} {\rm tr} R\wedge R\right)\,.
\end{equation}
Let us compute first the decay constant from the kinetic term: if we only have one axion, the only entry of the $\gamma $ matrix reads:
\begin{equation}
    \gamma=\int_X \beta\wedge\star \beta=\int_{C} \beta^2 \rm dVol_C \int_Y \rm dVol_Y=\mathcal{V}_C^{-1} \mathcal{V}_Y=\frac{\mathcal{V}}{\mathcal{V}_C^2}\,,
\end{equation}
since $\int_C \beta = \int_C \mathcal{V}_C^{-1} \rm dVol_C=1$.
From the kinetic energy term, 
we find
\begin{equation}
    f_b^2=\frac{\gamma}{2\pi g_s^2}= \frac{\mathcal{V}}{2\pi g_s^2 \mathcal{V}_C^2}\,.
\end{equation}
This setup is only a toy-model, however we can consider the case where the CY is a fibration with $Y=K3$ over $C=\mathbb{C P}^1$. This toy example cannot have $h^{1,1} <  2$, so to  qualitively mimic scenarios with a single model-dependent axion, we will consider only the axion arising from the base. Let us now look at the CS couplings of that axion. 
The instanton numbers $N_1,\, N_2$ for the two factors of the $E_8\times E_8$ gauge bundle over $Y$, defined as 
\begin{equation}
    N = \frac{1}{16\pi^2}\int {\rm tr} F \wedge F\,,
\end{equation}
are required to be $N_1,\,N_2\geq 0$ to satisfy SUSY constraints. The Bianchi identity requires $N_1+N_2=24$ since 
\begin{equation}
    \frac{1}{16\pi^2}\int_{K3} \left({\rm tr} R\wedge R - {\rm tr} F\wedge F\right)=\chi(K3) -\frac{1}{16\pi^2}\int_{K3} {\rm tr} F\wedge F= 24-\frac{1}{16\pi^2}\int_{K3} {\rm tr} F\wedge F=0\,,
\end{equation}
where $\chi(K3)=24$ is the Euler's characteristic.
Thus, from \cref{kprime}, 
\begin{equation}
    n=N_1-\frac{1}{2}\chi(K3)=N_1-12\quad \to |n|\leq 12\,.
\end{equation}

Let us now consider the two different cases: the absence of worldsheet instantons or the absence of gaugino condensation. 
In the absence of worldsheet instantons, the value for the mixing parameter reads \cref{alpha}
\begin{equation}
    \alpha=n \frac{6 g_s^2 }{v^2}\sim 0.1\,,
\end{equation}
Therefore, we need to go back to the canonical basis $\vartheta$, with decay constants $f_a$ and $f_1/n_1$, where the CS couplings read \cref{final_CS},
which, if taking $n=12$, $v\sim3$, and $g_s\sim 0.7$, become $ \lambda_{v}\sim 0.5$, $ \lambda_{h}\sim 1.1$. 
 
In the case of absence of gaugino condensation instead, we go to the mass basis $\varphi$, defined in \cref{phibasis}, where, taking the same values as before, the CS couplings read $\lambda_{1,v}\sim 0.5$, $\lambda_{1,h}\sim -0.47$, $\lambda_{2,h}\sim 0.03$. 

\subsubsection{Example: Quintic}
\label{sec:quintic}
To illustrate the general mechanism, we consider a more concrete example based on the quintic Calabi-Yau threefold $\mathbb{CP}^4[5]$ with $h^{1,1} = 1$. We define a two-axion model with one model-independent axion $\vartheta_a$ and one model-dependent axion $\vartheta_1$.

The quintic is reviewed in Appendix \ref{ap_5}. The triple intersection number reads $\kappa_{111}=5$, such that that the internal volume is given by
\begin{equation}
    \mathcal{V} = \frac{1}{6} \kappa_{111} v^3 = \frac{5}{6} v^3,
\end{equation}
where $v$ is the volume of the single 2-cycle. The decay constants are estimated as
\begin{align}
    f_a &= \frac{g_s}{\sqrt{2\pi \mathcal{V}}} = \sqrt{\frac{3}{5}} \cdot \frac{g_s}{\sqrt{\pi v^3}}, \\
    f_1 &= \frac{v}{\sqrt{2\pi} g_s}.
\end{align}

Let us now compute the Chern class of the Calabi-Yau, denoting $H$ as the hypersurface corresponding to the single divisor of the quintic: 
\begin{equation}
    \begin{aligned}
    c(T_X)&=c(T_A)/c(N_X)= \frac{(1+H)^5}{(1+5H)}=
    &=1+10H^2+..\\
    \end{aligned}
\end{equation}
Therefore $c_1(T_X)=0$, consistently with the CY condition, and $c_2(T_X)=10H^2$. 

Following~\cite{Blumenhagen:2005ga} we consider the gauge bundle $W=V_1+V_2+L$, where $V_1, V_2$ are SU(4) gauge bundles and $L$ is a line bundle. $V_1$ and $L $ are embedded in the hidden $E_8$ whereas $V_2$ is in the visible one.  The bundle $V_1$ is defined via the cohomology of the short exact sequence, known here as a monad,
\begin{equation}
    0 \longrightarrow \mathcal{O}\big|_X \xrightarrow{M} \mathcal{O}(1)^{\oplus 5} \oplus \mathcal{O}(3) \xrightarrow{N} \mathcal{O}(8) \longrightarrow 0 \, ,
\end{equation}
where $\mathcal{O}(1)^{\oplus 5} = \mathcal{O}(1) \oplus \dots \oplus \mathcal{O}(1)$.  
The bundle is then
\begin{equation}
    V = \frac{\ker(N)}{\mathrm{Im}(M)} \, .
\end{equation}
Its rank is
\begin{equation}
    \mathrm{rk}(V) = \mathrm{rk}\big(\mathcal{O}(1)^{\oplus 5} \oplus \mathcal{O}(3)\big) - \mathrm{rk}\big(\mathcal{O}|_X\big) - \mathrm{rk}\big(\mathcal{O}(8)\big) = 4 \, ,
\end{equation}
confirming that it is an $SU(4)$ gauge bundle.

The total Chern class is
\begin{equation}
\begin{aligned}
    c(V) &= \frac{c\big(\mathcal{O}(1)^{\oplus 5} \oplus \mathcal{O}(3)\big)}{c\big(\mathcal{O}|_X\big) \, c\big(\mathcal{O}(8)\big)} 
    = \frac{(1+H)^5 (1+3H)}{(1+8H)} \\
    &= (1+H)^5 (1+3H) \big(1 - 8H + 64H^2 + \dots\big) \\
    &= 1 + 25H^2 + \dots
\end{aligned}
\end{equation}
so that $c_2(V) = 25H^2$. We can now compute the topological CS coupling: 
\begin{equation}
\begin{aligned}
    n_1 &= \int_X \beta \wedge \frac{1}{16\pi^2} \left[ \mathrm{tr}_1(\overline{F} \wedge \overline{F}) - \frac12 \mathrm{tr}(\overline{R} \wedge \overline{R}) \right] \\
        &= \int_X \beta \wedge \left[ c_2(V_1) - \frac12 c_2(TX) \right] \\
        &= \left( 25 - \frac12 \cdot 10 \right) \int_X \beta \wedge H^2 \\
        &= 20 \int_X \beta \wedge H^2 \, .
\end{aligned}
\end{equation}
Since ${\rm PD}[\beta] = \Pi_4$ is a 4-cycle and $h^{1,1}(X) = 1$, every 4-cycle is homologous to $H$, so $[\Pi_4] = m[H]$.  
With the normalization $\int_{\Sigma} \beta = 1$, we set $m = 1$, hence
\begin{equation}
    n_1 = 20 \, H^3 = 20 \times 5 = 100 \, .
\end{equation}
This is the Chern-Simons coefficient coupling the axion to the gauge sector.

\paragraph{\boxed{\text{GC}}}
The squashing parameter defined in \cref{alpha} becomes:
\begin{equation}
    \alpha = n \cdot g_s^2 \cdot \frac{6}{5 v^2}.
\end{equation}
To remain within the perturbative regime $\mathcal{V} \lesssim 20$, we bound $v \lesssim 2.5$, leading to
\begin{equation}
    \alpha \gtrsim 40 g_s^2 \gtrsim 20 \quad \text{for} \quad g_s \sim 0.7.
\end{equation}
This implies large kinetic mixing:
\begin{equation}
    \mathcal{L}_{\text{kin}} = \frac{1}{2} (\partial \varphi_1)^2 + \frac{1}{2} (\partial \varphi_2)^2 + \partial \varphi_1 \partial \varphi_2.
\end{equation}
We then rotate to the mass basis
\begin{equation}
    \chi_1 = \varphi_1 + \varphi_2 = \vartheta_a, \qquad \chi_2 = \varphi_1 - \varphi_2 = \alpha \vartheta_1,
\end{equation}
with decay constants
\begin{equation}
    f_{\chi_1} = f_a, \qquad f_{\chi_2} = \frac{f_1}{\alpha}\,,
\end{equation}
effectively lowering the model dependent axion decay constant by a factor of $\alpha$. 
The Chern-Simons couplings in this basis are
\begin{align}
    \lambda_{\chi_1,v}=-\lambda_{\chi_2,v} &= 8 n g_s^2 \cdot \left( \frac{1}{\frac{5\pi v^3}{6} + g_s^2 v n} \right), \\
    \lambda_{\chi_1,h} =-\lambda_{\chi_2,h}&= -8 n g_s^2 \cdot \left( \frac{1}{\frac{5\pi v^3}{6} - g_s^2 v n} \right).
\end{align}
For example, taking plausible values $v = 3$, $n = 100$, and $g_s = 0.7$ gives
\begin{equation}
    \lambda_{\chi_1,v} \approx 1.8, \qquad \lambda_{\chi_1,h} \approx -5.2,
\end{equation}
showing a visible/hidden hierarchy induced by the large anomaly coefficient.

\paragraph{\boxed{\text{noGC}}}

Let us now consider the possibility of no gaugino condensation. 
In this case there will be one axion only coupled to the dark sector, and one coupled to both dark and visible sector. 
The CS couplings are defined in \cref{two_axions_CS}. In this case, by taking $k=1$ and $g_s=0.7$, we can also estimate the decay constants as 
\begin{equation}
    f_a= \frac{g_s}{\sqrt{2\pi \mathcal{V}}}= \sqrt{\frac{3}{5}}\frac{g_s}{\sqrt{\pi v^3}}\,,\quad \quad f_1=\frac{v}{\sqrt{2\pi} g_s }
\end{equation}
The Chern-Simons coupling read
\begin{equation}
 \lambda_{\varphi_1,v} \sim 1.8 ,\quad \lambda_{\varphi_1,h}\sim -0.1\,,\quad \lambda_{\varphi_2,h}\sim -4.7\,.
\end{equation}
One axion is coupled only to the hidden sector, whereas the first one is coupled to both sectors but mainly to the visible one.

\subsection{Three-Axion System}

We now extend our setup to include three axions: two model-dependent axions $\vartheta_1, \vartheta_2$ and one model-independent axion $\vartheta_a$, with respective decay constants $f_1$, $f_2$, and $f_a$. The Chern-Simons couplings take the form:
\begin{equation}
\mathcal{L}_{\text{CS}} = \frac{1}{16\pi^2} \left(\frac{\vartheta_a}{f_a} + \frac{n_1}{f_1} \vartheta_1 + \frac{n_2}{f_2} \vartheta_2\right) {\rm tr}_1 F \wedge F + \frac{1}{16\pi^2} \left(\frac{\vartheta_a}{f_a} - \frac{n_1}{f_1} \vartheta_1 - \frac{n_2}{f_2} \vartheta_2\right) {\rm tr}_2 F \wedge F.
\end{equation}
The resulting potential receives contributions from QCD, gaugino condensation, and worldsheet instantons:
\begin{equation}
\begin{aligned}
V_{\text{mass}} = &- \Lambda_{\text{QCD}}^4 \cos\left(\frac{\vartheta_a}{f_a} + \frac{n_1}{f_1} \vartheta_1 + \frac{n_2}{f_2} \vartheta_2 +\delta \right)
- \Lambda_{\text{gc}}^4 \cos\left(\frac{\vartheta_a}{f_a} - \frac{n_1}{f_1} \vartheta_1 - \frac{n_2}{f_2} \vartheta_2\right) \\
&- \Lambda_{\text{ws},1}^4 \cos\left(\frac{\vartheta_1}{f_1}\right) - \Lambda_{\text{ws},2}^4 \cos\left(\frac{\vartheta_2}{f_2}\right).
\end{aligned}
\end{equation}

Expanding the potential to second order gives the mass matrix:
\begin{equation}
\begin{pmatrix}
 \frac{\Lambda_{\text{gc}}^4 + \Lambda_{\text{QCD}}^4}{f_a^2} & \frac{n_1 (\Lambda_{\text{QCD}}^4 - \Lambda_{\text{gc}}^4)}{f_a f_1} & \frac{n_2 (\Lambda_{\text{QCD}}^4 - \Lambda_{\text{gc}}^4)}{f_a f_2} \\
 \frac{n_1 (\Lambda_{\text{QCD}}^4 - \Lambda_{\text{gc}}^4)}{f_a f_1} & \frac{n_1^2 (\Lambda_{\text{QCD}}^4 + \Lambda_{\text{gc}}^4) + \Lambda_{\text{ws},1}^4}{f_1^2} & \frac{n_1 n_2 (\Lambda_{\text{QCD}}^4 + \Lambda_{\text{gc}}^4)}{f_1 f_2} \\
 \frac{n_2 (\Lambda_{\text{QCD}}^4 - \Lambda_{\text{gc}}^4)}{f_a f_2} & \frac{n_1 n_2 (\Lambda_{\text{QCD}}^4 + \Lambda_{\text{gc}}^4)}{f_1 f_2} & \frac{n_2^2 (\Lambda_{\text{QCD}}^4 + \Lambda_{\text{gc}}^4) + \Lambda_{\text{ws},2}^4}{f_2^2}
\end{pmatrix}.
\end{equation}
We now analyze this system in different limiting cases.

\paragraph{\boxed{\text{noGC}} - Isotropic}

We begin with the case where gaugino condensation is absent, and the axion potential receives contributions only from QCD and worldsheet instantons.
In isotropic compactifications with $v \sim \mathcal{V}^{1/2} \sim 3$, the worldsheet instanton contribution can dominate: even if $W_0\sim 10^{-13}M_{Pl},$
\begin{equation}
    \Lambda_{\text{ws}}^4 \sim 10^{-22} M_{\text{Pl}}^4 \gg \Lambda_{\text{QCD}}^4 \sim 10^{-75} M_{\text{Pl}}^4.
\end{equation}
Assuming $\Lambda_{\text{ws},1} \approx \Lambda_{\text{ws},2} \gg \Lambda_{\text{QCD}}$, and $f_1\sim f_2$, one axion (aligned with QCD) which is mostly $\vartheta_a$ remains light, while the other two become heavy. 

The mass basis $\varphi$ in this case will be the original $\vartheta$ one, at first order, where the light axion will be mainly $\vartheta_a$ while the model dependent axions will get contributions mainly from worldsheet instantons:
\begin{align}
    \frac{\varphi_1}{f_{\varphi_1}}&=\frac{\vartheta_a}{f_a}\,,\qquad m_{\varphi_1}^2\simeq \frac{\Lambda _{\text{QCD}}^4}{f_a^2}\,, \\
        \frac{\varphi_2}{f_{\varphi_2}}&=\frac{\vartheta_1}{f_1}\,,\qquad m_{\varphi_2}^2 =\frac{\Lambda _{\text{ws}}^4}{f_1^2},\\
    \frac{\varphi_3}{f_{\varphi_3}}&=\frac{\vartheta_2}{f_2}\,,\qquad m_{\varphi_3}^2 = \frac{\Lambda _{\text{ws}}^4}{f_1^2}\,.
\end{align}
All axions couple to both visible and hidden sectors, and the CS couplings can be evaluated by separating the contribution from the 
\begin{equation}
\label{CS_tilde}
    \lambda_{\varphi_i, \text{hv}} =\pm \tilde{\lambda}_{\varphi_i} \cdot \frac{\pm 8 n_i k g_s^2}{\pi \mathcal{V} \pm g_s^2 v_i n_i},
\end{equation}
with leading-order expressions for the $\tilde{\lambda}_{\varphi_i}\sim 1$. In the absence of gaugino condensation, one QCD axion is light, while the others may be much heavier, especially in isotropic compactifications.

\paragraph{\boxed{\text{noGC}} - Anisotropic}
In anisotropic compactifications, one worldsheet instanton term can be exponentially suppressed, allowing a second axion to remain light. In this regime, we obtain a light axion $\varphi_1$ orthogonal to the heavier combinations lifted by dominant worldsheet effects.
Taking $v \sim 30$ and small remaining cycles, we can still ensure
\begin{equation}
    \mathcal{V} = \frac{1}{6} \kappa_{ijk} v^i v^j v^k < 25,
\end{equation}
so that only one worldsheet instanton is suppressed. We model this by setting 
\begin{equation}
    \Lambda_{\text{ws},2}^4 = \epsilon \Lambda_{\text{ws},1}^4, \qquad \text{with} \quad \epsilon \ll 1,
\end{equation}
with small $\epsilon$ parameter. The axion mass basis is:
\begin{equation}
    \begin{aligned}
        \varphi_1&=\frac{\vartheta _2}{f_2}-\frac{n_2  \vartheta _a}{f_2}\,,\\ 
       \varphi_2&=\frac{f_2^2 \vartheta _a+\vartheta _2 n_2 f_a^2}{n_2 f_a^2 f_2} \,,\\ 
         \varphi_3&=\frac{\vartheta_2}{f_2}+\frac{\frac{ \vartheta _1 \left(n_2^2 f_a^2+f_2^2\right) \left(n_1^2 \Lambda _{\text{gc}}^4-\Lambda _{\text{ws,1}}^4\right)}{f_2 n_1 f_a^2 \Lambda _{\text{ws,1}}^4}-\frac{f_2 \vartheta _a}{f_a^2}+\frac{f_2 \vartheta _1 \left(\frac{\Lambda _{\text{ws,1}}^4}{\Lambda _{\text{gc}}^4}+n_1^2\right)}{f_1^2 n_1}}{n_2} \,,\\ 
    \end{aligned}
\end{equation}
with masses
\begin{equation}
\begin{aligned}
m_{\varphi_1}^2 &= \frac{\epsilon  \Lambda _{\text{ws,1}}^4}{n_2^2 f_a^2+f_2^2} , \\
m_{\varphi_2}^2 &=\Lambda _{\text{QCD}}^4 \left(\frac{1}{f_a^2}+\frac{n_2^2}{f_2^2}\right), \\
m_{\varphi_3}^2 &= \frac{\Lambda _{\text{ws,1}}^4}{f_1^2} \,.
\end{aligned}
\end{equation}
Thus, the axion $\varphi_1$ becomes ultra-light, $\varphi_2$ is at the QCD scale, and $\varphi_3$ is heavy. This hierarchy is only achievable in anisotropic scenarios.
 The CS couplings again take the form \cref{CS_tilde} and the leading $\tilde{\lambda}_{\varphi_i}$ are given by a combination of the decay constants and the $n_i$. We report them in \cref{app_CS_couplings} as they are lengthy and their functional form is not instructive. In anisotropic compactifications, therefore, an ultralight axion with visible couplings can arise in the absence of gaugino condensation.

\paragraph{\boxed{\text{GC}} - Isotropic }
When gaugino condensation is present, its associated scale typically dominates over QCD and may compete with worldsheet instantons depending on the compactification and the rank of the condensing gauge group.
We first consider the isotropic limit. The scale of gaugino condensation is given by
\begin{equation}
    \Lambda_{\text{gc}}^4 \sim \mu^4 e^{-\frac{2\pi}{N g^2 }\mathcal{V}}\,.
\end{equation}
Assuming similar $\mu$ for $V_{\text{gc}}$ and $V_{\text{ws}}$, the hierarchy becomes in the isotropic case:
\begin{equation}
     \Lambda_{\text{ws}}^4\equiv \Lambda_{\text{ws,1}}^4\simeq\Lambda_{\text{ws,2}}^4  \gg \Lambda_{\text{gc}}^4 \gg \Lambda_{\text{QCD}}^4\,.
\end{equation}
As a result, no axion remains at the QCD scale, and observable axions are generically heavy. 
The axion mass basis is:
\begin{equation}
    \begin{aligned}
        \varphi_1&=-\frac{\vartheta_2}{f_2}-\frac{\vartheta _1 n_2}{f_1 n_1}\,,\\ 
       \varphi_2&=\frac{\vartheta_a \left(f_a^2 \left(\frac{\Lambda _{\text{ws}}^4}{\Lambda _{\text{gc}}^4}+n_1^2+n_2^2\right)-f_1^2\right)}{f_1 n_2 f_a^2}+\frac{\vartheta_2}{f_2}+\frac{\vartheta _1 n_1}{f_1n_2} \,,\\ 
         \varphi_3&=\frac{\vartheta_2}{f_2}+\frac{\vartheta _1 n_2}{f_1 n_1} \,,\\ 
    \end{aligned}
\end{equation}
with masses
\begin{equation}
\begin{aligned}
m_{\varphi_1}^2 &= \frac{ \Lambda_{\text{ws}}^4}{f_1^2} , \\
m_{\varphi_2}^2 &= \Lambda_{\text{gc}}^4\frac{1}{f_a^2} , \\
m_{\varphi_3}^2 &= \frac{n_1^2 \Lambda _{\text{gc}}^4+n_2^2 \Lambda _{\text{gc}}^4+\Lambda _{\text{ws}}^4}{f_1^2}  \,.
\end{aligned}
\end{equation}

\paragraph{\boxed{\text{GC}} - Anisotropic }
In this regime, one worldsheet instanton is exponentially suppressed, allowing one axion to remain light (at the QCD axion mass scale) despite the presence of gaugino condensation.
\begin{equation}
     \Lambda_{\text{ws,1}}^4 \gg \Lambda_{\text{gc}}^4 \gg \Lambda_{\text{QCD}}^4\gg \Lambda_{\text{ws,2}}\,.
\end{equation}

\begin{equation}
    \begin{aligned}
        \varphi_1&=\frac{\vartheta_2}{f_2}+\frac{\vartheta _a n_2}{f_2 }\,,\\
       \varphi_2&=-\frac{f_2^2 \vartheta _a-\vartheta _2 n_2 f_a^2}{n_2 f_a^2 f_2} \,,\\ 
         \varphi_3&=\frac{\vartheta_2}{f_2}+\frac{\frac{ \vartheta _1 \left(n_2^2 f_a^2+f_2^2\right) \left(n_1^2 \Lambda _{\text{gc}}^4-\Lambda _{\text{ws,1}}^4\right)}{f_2 n_1 f_a^2 \Lambda _{\text{ws,1}}^4}-\frac{f_2 \vartheta _a}{f_a^2}+\frac{f_2 \vartheta _1 \left(\frac{\Lambda _{\text{ws,1}}^4}{\Lambda _{\text{gc}}^4}+n_1^2\right)}{f_1^2 n_1}}{n_2} \,,\\ 
    \end{aligned}
\end{equation}
with masses
\begin{equation}
\begin{aligned}
m_{\varphi_1}^2 &= \frac{4 n_2^2 \Lambda _{\text{QCD}}^4}{n_2^2 f_a^2+f_2^2} , \\
m_{\varphi_2}^2 &=\Lambda _{\text{gc}}^4 \left(\frac{1}{f_a^2}+\frac{n_2^2}{f_2^2}\right) , \\
m_{\varphi_3}^2 &= \frac{n_1^2 \Lambda _{\text{gc}}^4+\Lambda _{\text{ws,1}}^4}{f_1^2}  \,.
\end{aligned}
\end{equation}

\subsubsection{Example: Bi-cubic CICY}
\label{sec:bicubic}
Let us consider the bi-cubic CICY defined as a degree-(3,3) hypersurface $\mathbb{P}^2\times \mathbb{P}^2$ with Hodge numbers $(h^{1,1},h^{2,1})=(2,83)$. 
\begin{equation}
\begin{array}{c@{\;}c}
\begin{array}{c}
\mathbb{P}^2 \\
\mathbb{P}^2
\end{array}\left[
\begin{array}{c}
3 \\
3
\end{array}
\right] \,,
\end{array}
\end{equation}
Its Chern class and intersection numbers are computed in Appendix \ref{ap_bicubic}
Following~\cite{Anderson:2009mh}, we consider a vector bundle $V$ defined as 
\begin{equation}
   0 \rightarrow V \rightarrow  \mathcal{O}(1,0)^{\oplus 3} \oplus \mathcal{O}(0,1)^{\oplus 3} \rightarrow \mathcal{O}(1,1) \oplus \mathcal{O}(2,2) \rightarrow 0
\end{equation}
which gives a bundle with structure group $G=SU(4)$, such that in 4D the gauge group is a $SO(10)$ GUT, which reproduces an MSSM-like spectrum after a suitable Wilson line breaks the $SO(10)\to SU(3)\times SU(2)\times U(1)_Y\times U(1)_{B-L}$.
In the original example in order to satisfy anomaly cancellation they considered $M5$-branes, such that the hidden sector bundle $\tilde{V}$ could remain trivial. If $c_2(TX)-c_2(V)$ is an effective class on the CY, then the anomaly and the effectiveness conditions are automatically satisfied for a trivial hidden bundle and a five brane class $W= c_2(TX)-c_2(V)$~\cite{Anderson:2008uw}, as the actual anomaly cancellation condition reads 
\begin{equation}
    c_2(TX)-c_2(V)-c_2(\tilde{V})=[W]\,,
\end{equation}
where $[W]$ in an effective five-brane class. 
However, we are interested in the hidden gauge sector, and therefore we chose a non-Abelian vector bundle with structure group $SU(N)$ in the hidden sector, such that the resulting 4D gauge group is its commutant inside $E_8$.
Using the basis of divisors $H_1,\, H_2$ corresponding to the two $\mathbb{P}^2$ factors, the second Chern class of the tangent bundle reads 
\begin{equation}
    c(TX)=\frac{(1+H_1)^3(1+H_2)^3)}{(1+3H_1+3H_2)}\qquad c_2(TX)= 3H_1^2+3H_2^2+9H_1H_2\,,
\end{equation}
while the second Chern class of the monad bundle reads 
\begin{equation}
    c(V)=\frac{(1+H_1)^3(1+H_2)^3)}{(1+H_1+H_2)(1+2H_1+2H_2)}\qquad c_2(V)= H_1^2+H_2^2+5H_1H_2\,.
\end{equation}
To satisfy the anomaly cancellation condition, taking $[W]=0$, we need 
\begin{equation}
\label{hiddengauge}
    c_2(\tilde{V})=2H_1^2+2H_2^2+4H_1H_2 \qquad c_1(\tilde{V})=0\,.
\end{equation}
Take 
\begin{equation}
    0 \longrightarrow \tilde{V} \longrightarrow \bigoplus_i^{r+s} \mathcal{O}\left(a_i, b_i\right) \xrightarrow{f} \bigoplus_{i=1}^s \mathcal{O}\left(c_j, d_j\right) \longrightarrow 0
\end{equation}
and ask to satisfy the conditions \cref{hiddengauge}.
This translates into 
\begin{equation}
\begin{aligned}
    \sum_i a_i=\sum_j c_j \qquad &\sum_i b_i=\sum_j d_j,\\
    \sum_i \frac12\left(a_i^2 H_1^2 +2a_i b_i H_1H_2+b_i^2H_2^2\right)&=\sum_j\frac{1}{2}\left(c_j^2H_1^2 +2 c_jd_jH_1H_2+d_j^2H_2^2\right)\,.
\end{aligned}
\end{equation}
One example of monad construction that satisfies this is 
\begin{equation}
     0 \longrightarrow \tilde{V} \longrightarrow \mathcal{O}\left(1,0\right)^{\oplus 6}\oplus\mathcal{O}\left(0,1\right)^{\oplus 4} \xrightarrow{f}  \mathcal{O}\left(2,0\right)^{\oplus 2}\oplus \mathcal{O}\left(1,2\right)^{\oplus 2} \longrightarrow 0
\end{equation}
which is a rank $6$ gauge bundle giving an $SU(2)$ hidden gauge sector in 4D. This can be further broke down to abelian $U(1)$s via Wilson lines. 
The tree level DUY equation is satisfied, as $c_1(V)=0$.
The bundle's stability is assured if any subsheaf $\mathcal{F}\in V$ with $0<\mathrm{rk}(\mathcal{F})<\mathrm{rk}(V)$ has $\mu(\mathcal{F})<\mu(V)=0$. Therefore, one would need to check that for every subsheaf the slope is negative. One way to ensure this is to check that $H^0(X,V)=0$, which would be sufficient~\cite{Anderson:2008uw}. 

Another possibility is instead to have the line bundle $\mathcal{L}=\mathcal{O}(2,2)$, for which the second Chern character reads $\mathrm{ch}_2(\mathcal{L})=\frac12 c_1^2=\frac{1}{2}\left(2H_1+2H_2\right)^2$. This would result in a 4D gauge group that looks like $E_7\times U(1)$, where the $U(1)$ would be anomalous as it is there also in the structure group, and thus becomes massive by eating one of the two model dependent axions. This will always happen when there is a line bundle in the structure group. In this case we are effectively back to the two-axion scenario, with 
\begin{equation}
    n_1=n_2=\int H_1 \left(c_2(V)-\frac12c_2(TX)\right)=3\,.
\end{equation}

\subsubsection{Example: CICY with $U(4)$ Bundle}

Let us take a $U(4)$ bundle on the CICY studied in~\cite{Blumenhagen:2005ga}
\begin{equation}
\begin{array}{c@{\;}c}
\begin{array}{c}
\mathbb{P}^3 \\
\mathbb{P}^1
\end{array}\left[
\begin{array}{c}
4 \\
2
\end{array}
\right] \,,
\end{array}
\end{equation}
which has $h^{1,1}=2$ and $h^{2,1}=86$.
Calling $\eta_1$ the 2-form defined on $\mathbb{P}_3$ and $\eta_2$ the two-form defined on $\mathbb{P}_1$, the Stanley-Reissner ideal, which contains those coordinates that cannot be set to zero simultaneously (or equivalently, those divisors which do not intersect) can be read from the D-terms to be
\begin{equation}
    SR=\{\eta_1^4,\eta_2^2\}\,.
\end{equation}
The intersection form reads 
\begin{equation}
    I_3=2\eta_1^3+4\eta_1^2\eta_2\,.
\end{equation}
Therefore, there exist 2 possible 4-forms on the CY:
\begin{equation}
    \{\eta_1^2,\eta_1\eta_2\}\,.
\end{equation}
The Chern classes can be computed via 
\begin{equation}
    c(TX)=\frac{(1+\eta_1)^2(1+\eta_2^2)}{(1+4\eta_1+2\eta_2)}=...= 1+6\eta_1^2+8 \eta_1\eta_2+...
\end{equation}
The second Chern class can be read from the equation above to be $c_2(TX)=6\eta_1^2+8\eta_1\eta_2$.
The gauge bundle chosen to reproduce the SM-like sector in the first $E_8$ reads 
\begin{equation}
    W=V\oplus L^{-1}
\end{equation}
where the line bundle is taken to be $L=\mathcal{O}(-2,2)$ and the $U(4)$ bundle $V$ is defined via 
\begin{equation}
    \left.\left.0 \rightarrow V \rightarrow \mathcal{O}(1,0)^{\oplus 2} \oplus \mathcal{O}(0,1)^{\oplus 2} \oplus \mathcal{O}(1,1)^{\oplus 2}\right|_{CY} \xrightarrow{f} \mathcal{O}(4,1) \oplus \mathcal{O}(2,1)\right|_{CY} \rightarrow 0
\end{equation}
such that it satisfies the tadpole condition
\begin{equation}
    c_2(V)-c_1^2(L)=c_2(TX)\,,
\end{equation}
and the map $f$ is chosen to not degenerate at any point. 
One can indeed check that 
\begin{equation}
    c(V)=\frac{(1+\eta_1)^2(1+\eta_2^2)(1+\eta_1+\eta_2)^2}{(1+4\eta_1+\eta_2)(1+2\eta_1+\eta_2)}\,,
\end{equation}
such that 
\begin{equation}
    c_2(V)=\eta_2^2+10\eta_1^2=10\eta_1^2\,,
\end{equation}
since $\eta_2^2=0$ as for the SR. 
Taking the two axions defined as the dimensional reduction of the $B_2$ as 
\begin{equation}
B=b_1 \eta_1+b_2\eta_2 \,,   
\end{equation}
we can compute the $n_i$ as 
\begin{equation}
\begin{aligned}
  n_1 &=  \int \eta_1\wedge  \left(c_2(V)-c_1^2(L)-\frac12 c_2(TX)\right)=\int \eta_1\wedge  \left(\frac12 c_2(TX)\right)\\
  &= 3 \int \eta_1^3 +4 \int \eta_1^2\eta_2= 6+16=22\\
n_2&=  \int \eta_2\wedge\left(\frac12 c_2(TX)\right)= 3 \int \eta_1^2\eta_2= 6\,. 
\end{aligned}
\end{equation}
Since $W$ was taken such that there is no gauge bundle embedded in the hidden sector, the 4D gauge group remains E$_8$.

\paragraph{\boxed{\text{noGC}}}
To arrive at cases where the hidden $E_8$ is fully broken down to just $U(1)$-factors, we may need to turn on non-trivial Wilson lines if the gauge bundle is insufficient to do the  full breaking on its own. Getting such Wilson lines requires the CY to have a non-trivial first homotopy group. Besides a very small number directly existing within known sets of CYs such as the CICYs, such CY manifolds can be obtained by modding out a freely acting discrete involution from an original CY possessing the required discrete symmetry~\cite{Anderson:2018heq,Anderson:2019kmx,Gray:2021kax}.

Given that we only have two non vanishing intersection numbers, $\kappa_{111}$ and $\kappa_{112}$, we find that the volume reads %
\begin{equation}
    \mathcal{V}=\frac16\left(2 v_1^3 + 4 v_1^2 v_2\right)\,.
\end{equation}
In this case, let us look at the possible CS couplings. 
In the isotropic case, we take $v_1\sim v_2\sim 3$, we find that the CS couplings read 
\begin{equation}
\begin{aligned}
    \lambda_{\varphi_1,v}&\sim 2.5 \,,\quad \lambda_{\varphi_2,v}\sim 1.2 \,,\quad \lambda_{\varphi_3,v}\sim -0.02\\
    \lambda_{\varphi_1,h}&\sim 1.4 \,,\quad \lambda_{\varphi_2,h}\sim 0.7 \,,\quad \lambda_{\varphi_3,v}\sim 0.01
\end{aligned}
\end{equation}

In this example it is difficult to construct an anisotropic case: knowing the intersection numbers from the expression of the volume, we see that if we take the limiting case $v_1\sim 1$, in order to maintain the volume smaller than $\sim 20$, we have to take at most $v_2\sim 5$. In this case we find
\begin{equation}
\begin{aligned}
    \lambda_{\varphi_1,v}&\sim 8.2 \,,\quad \lambda_{\varphi_2,v}\sim 6.1 \,,\quad \lambda_{\varphi_3,v}\sim 0.5\\
    \lambda_{\varphi_1,h}&\sim 29.4 \,,\quad \lambda_{\varphi_2,h}\sim 40.1 \,,\quad \lambda_{\varphi_3,h}\sim -2.4\,.
\end{aligned}
\end{equation}

\subsection{Three Axion Summary}
Let us summarize our findings by recalling that $\epsilon$ is a small parameter used in the anisotropic case relating the two worldsheet instanton scales, and the hierarchy: 
\begin{equation}
        \Lambda_{\text{ws}}^4 \gg \Lambda_{\text{gc}}^4 \gg \Lambda_{\text{QCD}}^4\gg \epsilon \Lambda_{\text{ws}}^4\,.
\end{equation}

\begin{table}[h!]
    \centering
    \renewcommand{\arraystretch}{2}
    \setlength{\tabcolsep}{10pt}
    \begin{tabular}{cc}
          \begin{tabular}{cc} 
          \textbf{\boxed{\text{noGC}}\, Isotropic} & \textbf{FDM}\\
             $m_{\varphi}^2=\left(  \begin{array}{c}
                  \frac{\Lambda _{\text{ws}}^4}{f_1^2}\\
                  \frac{\Lambda _{\text{QCD}}^4}{f_a^2}\\
                  \frac{\Lambda _{\text{ws}}^4}{f_1^2}
            \end{array}\right)$ & $\begin{array}{c}
                  \times
                  \end{array}$
            \end{tabular}
 
        &
        \begin{tabular}{cc}
            \textbf{\boxed{\text{noGC}}\, Anisotropic} &\textbf{FDM}\\
              $m_{\varphi}^2=\left(  \begin{array}{c}
              \epsilon 
                  \frac{ \Lambda_{\text{ws}}^4}{n_2^2 f_a^2+f_2^2}  \\
                  \Lambda _{\text{QCD}}^4 \left(\frac{1}{f_a^2}+\frac{n_2^2}{f_2^2}\right) \\
                  \frac{\Lambda _{\text{ws}}^4}{f_1^2}  \end{array}\right)$\quad &$\begin{array}{c}
                  \checkmark
                  \end{array}$
            \end{tabular}
        \\
        \begin{tabular}{cc} 
          \textbf{ \boxed{\text{GC}}\, Isotropic} & \textbf{FDM}\\
             $m_{\varphi}^2=\left(\begin{array}{c}
             \frac{ \Lambda_{ws}}{f_1^2}  \\
             \frac{\Lambda_{gc}^4}{f_a^2}  \\
             \frac{\Lambda _{\text{ws}}^4}{f_1^2}
            \end{array}\right)$ & $\begin{array}{c}
                  \times
                  \end{array}$
            \end{tabular}
        &
        \begin{tabular}{cc} 
          \textbf{\boxed{\text{GC}}\, Anisotropic} & \textbf{FDM}\\
             $m_{\varphi}^2=\left(\begin{array}{c}
             \frac{4 n_2^2 \Lambda _{\text{QCD}}^4}{n_2^2 f_a^2+f_2^2} \\
             \Lambda _{\text{gc}}^4 \left(\frac{1}{f_a^2}+\frac{n_2^2}{f_2^2}\right)  \\
             \frac{n_1^2 \Lambda _{\text{gc}}^4+\Lambda _{\text{ws}}^4}{f_1^2}
            \end{array}\right)$ & $\begin{array}{c}
                  \times
                  \end{array}$
            \end{tabular}
    \end{tabular}
    \caption{Summary of Mass Matrices in Different Regimes. The last column refers to the possibility of having a fuzzy dark matter candidate. This is available only in the \boxed{\text{noGC}} anisotropic case, where the FDM candidate aligns with $\varphi_1$.  }
    \label{tab:upshot2}
\end{table}

We find that a fuzzy dark matter candidate can arise only in the \boxed{\text{noGC}} anisotropic configuration. In all other setups, the hierarchy of the contirbutions to the scalar potential prevents the presence of an extremely light axion.

Our analysis covered both two- and three-axion systems. A compactification with larger $h^{1,1}$, and thus a greater number of axions, would proceed analogously, so higher-$h^{1,1}$ cases are not discussed here as they would not be illustrative. The general conclusion is that at least one potential term must be absent (as in the \boxed{\text{noGC}} case) or strongly suppressed (as in the anisotropic case) to address the Strong CP problem. Achieving a fuzzy dark matter candidate requires both the absence of gaugino condensation and the presence of a highly suppressed worldsheet instanton, which requires a fibred CY.

The relevant physical description is given in the mass basis, where the Chern–Simons couplings become nontrivial combinations of the decay constants and topological quantities. In certain compactifications, it is possible to achieve a clean separation between couplings to the visible and hidden sectors. In special cases, some couplings vanish entirely, allowing for axions that interact exclusively with one of the two sectors. Such configurations must, however, be examined on a case-by-case basis.


\section{Conclusions}
\label{sec:conclusions}

In this work, we have started the study of central aspects of the heterotic axiverse building on the foundations for axions in heterotic string theory with a focus on the mass spectrum and couplings of both the model-independent and model-dependent axions. In addition, we clarified when the strong–CP problem is solved or left unresolved.

Starting from the effective four-dimensional theory, we analyzed the kinetic structure, Chern–Simons couplings, and non-perturbative potentials generated by gauge and stringy instantons, including gaugino condensation and worldsheet effects.
We examined how axions acquire masses through these non-perturbative effects and under what conditions one linear combination remains sufficiently light and dominantly aligned with the QCD direction to solve the Strong CP problem. Particular attention was given to the alignment of non-perturbative terms and the role of kinetic mixing, showing that successful axion phenomenology in string compactifications depends not only on the presence of instanton corrections, but also on their relative alignment in axion field space. These constraints impose nontrivial requirements on the geometry and gauge bundle data of the compactification.
We illustrated these features with explicit heterotic construction on Calabi-Yau manifolds with $h^{1,1}=1,2$. These example highlight how decay constant hierarchies and physical axion couplings can be engineered in principle, but also emphasizes that achieving a light axion typically requires some care. Upon diagonalizing the mass and kinetic matrices, we extracted the physical decay constants and recast the Chern–Simons couplings in the mass basis, identifying the surviving light states and their coupling structure.

Altogether, our results show that the heterotic axiverse provides a compelling and highly constrained setting for axion phenomenology. While the presence of multiple axions is generic, realizing light axions, particularly those that can solve the Strong CP problem or play a role in cosmology, is not automatic. This observation has important implications for the landscape of viable string models with axionic dark matter or observable axion couplings. A look at \cref{tab:upshot} \& \cref{tab:upshot2} displays the outcome which arises as a consequence of combining the structure of the non-perturbative quantum effects providing the axion mass with the QCD instanton contribution and the constraints from maintaining CP quality as well as the rather strict upper bound on the compactification volume imposed by heterotic perturbativity. The resulting axion mass spectrum is quite different compared to the type IIB or M-theory axiverses -- most of the 2-form axions here stay rather heavy, while typically the axion responsible for solving the QCD CP problem (if possible) is the lightest axion state with a mass scale proportional to $\Lambda_{\text{QCD}}^4$. This has potential implications for the inflationary production of axions. For example, if inflation was driven by a K\"ahler modulus, then the two-form axions would be produced by parametric resonance similar to the discussion in~\cite{Leedom:2024qgr,Ling:2025nlw} where the dynamics relevant for heterotic axions is in the heavy axion regime described in~\cite{Leedom:2024qgr}. 
The only exception arises for fibred CY compactifications which break the hidden $E_8$ gauge group completely via gauge bundle choice and/or Wilson lines, and with their two dominant K\"ahler moduli stabilized in a highly anisotropic regime. For this rather special case, a suppressed world-sheet instanton direction can arise providing a single fuzzy-dark matter (FDM) candidate among the heterotic string axions.

Based on our results, several natural future directions of work now suggest themselves. First, it would be valuable to construct an explicit compactification that simultaneously (i) yields a realistic Standard Model sector, (ii) solves Strong CP with a QCD–aligned light axion, and (iii) hosts a hidden-sector multi-axion spectator dynamics capable of sourcing a gravitational-wave signal~\cite{Anber:2009ua,Dimastrogiovanni:2012ew,Dimastrogiovanni:2023juq,DAmico:2021fhz,DAmico:2021vka,Peloso:2016gqs,Namba:2015gja} or a spectral distortion signal~\cite{Kite:2020uix,Putti:2024uyr}. 
Concretely, this calls for a geometry–bundle–CS coupling pipeline. On the geometry side, one should pick a Calabi–Yau threefold with $h^{1,1}\!\ge\!2$ (to permit a nontrivial axion basis and kinetic mixings) and a polystable holomorphic vector bundle satisfying both the DUY equations and the heterotic Bianchi identity. The visible bundle data must engineer a SM-like gauge sector with massless hypercharge (no Green–Schwarz St\"uckelberg mass) and a chiral spectrum. The hidden bundle/gauge factors provide the non-perturbative dynamics (e.g. gaugino condensation) that generate axion masses and the gauge fields that couple to spectator axions.
On the axion side, one should compute the kinetic matrix from the Kähler potential, the charge/St\"uckelberg matrix for anomalous $\mathrm{U}(1)$’s to identify eaten directions, the anomaly coefficients governing Chern–Simons couplings, and the non-perturbative superpotential terms (gauge and worldsheet instantons) that lift axion combinations. Diagonalizing the kinetic and mass matrices then reveals whether a QCD-aligned light eigenstate exists with the required decay constant and domain-wall number, and how strongly the remaining spectators couple to hidden gauge fields.


\newpage

\acknowledgments
We thank Florent Baume, Craig Lawrie, Nicole Righi, and Fabian Ruehle for useful discussions. 
This article is based in part upon work from COST Action COSMIC WISPers CA21106, supported by COST (European Cooperation in Science and Technology). MP is supported by the Deutsche Forschungsgemeinschaft under Germany’s Excellence Strategy - EXC 2121 “Quantum Universe” - 390833306. AW is partially supported by the Deutsche Forschungsgemeinschaft under Germany’s Excellence Strategy - EXC 2121 “Quantum Universe” - 390833306 and by the Deutsche Forschungsgemeinschaft through the Collaborative Research Center SFB1624 ``Higher Structures, Moduli Spaces, and Integrability’'. JML was funded by the European Union and supported by the Czech Ministry of
Education, Youth and Sports (Project No. FORTE – CZ.02.01.01/00/22\_008/0004632).

\appendix

\section{Quintic Calabi-Yau}
\label{ap_5}

In this appendix we illustrate the simplest example of a Calabi-Yau threefold used in \cref{sec:quintic}: the quintic hypersurface
$\mathbb{CP}^4[5]$, defined by 
\begin{equation}
    \sum_{i=1}^5 z_i^5 = 0
\end{equation}
in $\mathbb{CP}^4$, with homogeneous coordinates $z_i$ and divisors $D_i = \{ z_i = 0 \}$.  
All $D_i$ are linearly equivalent, so we may identify the hyperplane class $H \equiv D_i$ for all $i$.

\subsubsection*{Geometry of the quintic}

The geometric $\tilde{F}$-term encodes the hypersurface condition, while the $\tilde{D}$-term describes the Kähler quotient:
\begin{equation}
  \tilde{F}:\quad z_1^5 + z_2^5 + z_3^5 + z_4^5 + z_5^5 = 0 \, , 
  \qquad
  \tilde{D}:\quad |z_1|^2 + \dots + |z_5|^2 = b \, ,
\end{equation}
with $b \geq 0$ for $X$ to be in the Kähler cone.

Since all $Z_a$ have the same gauge charge, the divisors $D_i$ are equivalent and generate $H^{1,1}(X)$:
\begin{equation}
    \mathcal{D} \equiv H \, , \quad h^{1,1}(X) = 1 \, .
\end{equation}

\subsubsection*{Triple intersection number}

To compute $H^3$, set $z_1 = z_2 = z_3 = 0$. The $\tilde{F}$-term reduces to
\begin{equation}
    z_4^5 + z_5^5 = 0 \, .
\end{equation}
This equation in $\mathbb{CP}^1$ has 5 distinct solutions for the ratio $z_4/z_5$:
\begin{equation}
    \frac{z_4}{z_5} = e^{i(2k+1)\pi/5} \, , \quad k = 0,1,2,3,4 \, .
\end{equation}
Thus, the three divisors intersect in five points, and we find
\begin{equation}
    H^3 = 5 \, .
\end{equation}

\subsubsection*{Tangent bundle Chern classes}

The total Chern class of the tangent bundle is
\begin{equation}
    c(T_X) = \frac{c(T_{\mathbb{CP}^4})}{c(N_X)} = \frac{(1+H)^5}{(1+5H)} \, .
\end{equation}
Expanding the denominator as $(1+5H)^{-1} = 1 - 5H + 25H^2 + \dots$ and multiplying out, we find
\begin{equation}
    c(T_X) = 1 + 10H^2 + \dots
\end{equation}
so that $c_1(T_X) = 0$ (as required for a Calabi--Yau) and $c_2(T_X) = 10H^2$.

\section{Bi-cubic CICY}
\label{ap_bicubic}
Consider the bi-cubic complete intersection Calabi-Yau defined as a degree-$(3,3)$ hypersurface in $\mathbb{P}^2_x \times \mathbb{P}^2_y$, with Hodge numbers $(h^{1,1},h^{2,1})=(2,83)$, used in the example \cref{sec:bicubic}:
\begin{equation}
\begin{array}{c@{\;}c}
\begin{array}{c}
\mathbb{P}^2 \\
\mathbb{P}^2
\end{array}\left[
\begin{array}{c}
3 \\
3
\end{array}
\right] \,.
\end{array}
\end{equation}

Let $h_1$ and $h_2$ denote the hyperplane classes of the two $\mathbb{P}^2$ factors, pulled back to the ambient space $A = \mathbb{P}^2_x \times \mathbb{P}^2_y$, so that
\begin{equation}
    h^3 = 0 \, , \quad k^3 = 0 \, , \quad \int_A h_1^2 h_2^2 = 1 \, .
\end{equation}
The bicubic hypersurface $X$ has class
\begin{equation}
    [X] = 3h_1 + 3h_2 \, .
\end{equation}

We take the basis of divisors on $X$ to be $H_1 = h_1|_X$ and $H_2 = h_2|_X$.  
The triple intersection numbers are then
\begin{equation}
\begin{aligned}
    \kappa_{111} &= \int_A h_1^3 \wedge (3h_1+3h_2) = 0 \, , \\
    \kappa_{222} &= \int_A h_2^3 \wedge (3h_1+3h_2) = 0 \, , \\
    \kappa_{112} &= \int_A h_1^2 k \wedge (3h_1+3h_2) = 3 \int_A h_1^2 h_2^2 = 3 \, , \\
    \kappa_{122} &= \int_A h_1 h_2^2 \wedge (3h_1+3h_2) = 3 \int_A h_1^2 h_2^2 = 3 \, .
\end{aligned}
\end{equation}
By symmetry of $\kappa_{abc}$, we have $\kappa_{121} = \kappa_{211} = 3$ and $\kappa_{212} = \kappa_{221} = 3$.

The intersection polynomial is therefore
\begin{equation}
    \mathcal{I}(v_1,v_2) = \sum_{a,b,c} \kappa_{abc} \, v_a v_b v_c
    = 3\,v_1^2 v_2 + 3\,v_1 v_2^2
    = 3\,v_1 v_2 \,(v_1 + v_2) \, .
\end{equation}

\paragraph{Volume }
With Kähler form $J = v_1 H_1 + v_2 H_2$, the Calabi-Yau volume is
\begin{equation}
    \mathcal{V} = \frac{1}{6} \int_X J^3
    = \frac{1}{6} \left( 3\,v_1^2 v_2 + 3\,v_1 v_2^2 \right)
    = \frac{1}{2} \left( v_1^2 v_2 + v_1 v_2^2 \right) .
\end{equation}

\paragraph{Chern classes}
Let $A=\mathbb{P}^2_x\times\mathbb{P}^2_y$ with hyperplane classes $h_1,h_2$ pulled back from the two factors.
For $A$ one has $c(T_A)=c(T_{\mathbb{P}^2_x})\,c(T_{\mathbb{P}^2_y})=(1+h_1)^3(1+h_2)^3$.
The bi-cubic hypersurface $X\subset A$ has class $[X]=3h_1+3h_2$.
By adjunction,
\begin{equation}
  c(T_X) = \frac{(1+h_1)^3(1+h_2)^3}{1+3h_1+3h_2}\bigg|_X= \frac{(1+H_1)^3(1+H_2)^3}{1+3H_1+3H_2} \, .
\end{equation}
Expanding to second order, we find
\begin{equation}
  c(T_X) 1+ \big(3H_1^2 + 9H_1H_2 + 3H_2^2\big) + \cdots \, .
\end{equation}
Hence
\begin{equation}
  c_1(T_X)=0 \, , \qquad c_2(T_X) = 3H_1^2 + 9H_1 H_2 + 3H_2^2 = 3\,(H_1^2 + 3\,H_1H_2 + H_2^2) \, ,
\end{equation}

\section{CS couplings}
\label{app_CS_couplings}
We report the $\tilde{\lambda}_i$ defined in the text at \cref{CS_tilde} for the \boxed{\text{noGC}} - anisotropic case: 
\begin{align}
    \tilde{\lambda}_{\varphi_1,v} &\sim \frac{f_1^4 n_2^3 (-f_a) + f_2 f_1^2 n_1 n_2^2 f_a (f_1 - n_1 f_a) + f_2^3 n_1^2 (2 f_1 n_1 f_a - n_1^2 f_a^2 + f_1^2)}{2 f_1^2 f_2 n_1 n_2 f_a \sqrt{f_2^2 n_1^2 + f_1^2 n_2^2}}, \\
    \tilde{\lambda}_{\varphi_2,v} &\sim \frac{f_1^4 n_2^3 f_a + f_2 f_1^2 n_1 n_2^2 f_a (f_1 - n_1 f_a) - f_2^3 n_1^2 (n_1^2 f_a^2 + f_1^2)}{2 f_1^2 f_2 n_1 n_2 f_a \sqrt{f_2^2 n_1^2 + f_1^2 n_2^2}}, \\
    \tilde{\lambda}_{\varphi_3,v} &\sim \frac{f_2 n_1 f_a + f_1 (n_2 f_a + f_2)}{f_a \sqrt{f_2^2 n_1^2 + f_1^2 n_2^2}},
\end{align}
\begin{align}
    \tilde{\lambda}_{\varphi_1,h} &\sim \frac{f_1^4 n_2^3 f_a - f_2 f_1^3 n_1 n_2^2 f_a + f_2 f_1^2 n_1^2 (f_2^2 - n_2^2 f_a^2) - 2 f_2^3 f_1 n_1^3 f_a - f_2^3 n_1^4 f_a^2}{2 f_1^2 f_2 n_1 n_2 f_a \sqrt{f_2^2 n_1^2 + f_1^2 n_2^2}}, \\
    \tilde{\lambda}_{\varphi_2,h} &\sim \frac{-f_1^4 n_2^3 f_a - f_2 f_1^3 n_1 n_2^2 f_a - f_2 f_1^2 n_1^2 (n_2^2 f_a^2 + f_2^2) - f_2^3 n_1^4 f_a^2}{2 f_1^2 f_2 n_1 n_2 f_a \sqrt{f_2^2 n_1^2 + f_1^2 n_2^2}}, \\
    \tilde{\lambda}_{\varphi_3,h} &\sim \frac{f_1 (f_2 - n_2 f_a) - f_2 n_1 f_a}{f_a \sqrt{f_2^2 n_1^2 + f_1^2 n_2^2}}.
\end{align}
%


\bibliographystyle{JHEP}
\bibliography{Het}
\end{document}